\documentclass[a4paper,11pt]{article}
\pdfoutput=1\usepackage{jheppub}
\usepackage{amsmath,amssymb,euscript}
\usepackage{slashed}
\usepackage{xspace}
\usepackage{color}
\usepackage{hyperref}
\usepackage{epsfig}
\usepackage{xcolor}
\usepackage{xspace}
\usepackage{verbatim}
\usepackage{multirow}
\usepackage{booktabs,graphicx}
\usepackage{mathtools}
\usepackage{tabulary}
\usepackage{soul} 
\usepackage{afterpage}
\newcolumntype{K}[1]{>{\centering\arraybackslash}p{#1}}

\graphicspath{{figs/}}
\emergencystretch=\maxdimen
\newcommand{\Gamth}{\text{$\langle \Gamma\rangle$}}

\begin{document}
\title{Inelastic Dirac Dark Matter}

\author[1]{Anastasiia Filimonova,}
\author[2,3,4]{Sam Junius,}
\author[2,3]{Laura Lopez Honorez,}
\author[1,5,6]{Susanne Westhoff\,}

\affiliation[1]{
\href{https://www.nikhef.nl/en/}{\color{black} Nikhef},
Science Park 105, 1098 XG Amsterdam, The Netherlands}
\affiliation[2]{Service de Physique Th\'eorique, Universit\'e Libre de Bruxelles, C.P. 225, B-1050 Brussels, Belgium}
\affiliation[3]{Theoretische Natuurkunde \& The International Solvay Institutes, \\
	Vrije Universiteit Brussel, Pleinlaan 2, B-1050 Brussels, Belgium}
\affiliation[4]{Inter-University Institute for High Energies, Vrije Universiteit Brussel, \\
Pleinlaan 2, B-1050 Brussels, Belgium}
\affiliation[5]{Institute for Theoretical Physics, Heidelberg University, 69120 Heidelberg, Germany}
\affiliation[6]{Institute for Mathematics, Astrophysics and Particle Physics, Radboud University,
Heyendaalseweg 135, 6525 AJ Nijmegen, The Netherlands}

\emailAdd{a.filimonova@nikhef.nl}
\emailAdd{sam.junius@vub.be}
\emailAdd{llopezho@ulb.be}
\emailAdd{susanne.westhoff@ru.nl}

		\begin{flushright}
		  P3H-22-006\\
                  ULB-TH/22-01\\
						                  Nikhef 2022-002
		\end{flushright}

\abstract{Feebly interacting thermal relics are promising dark matter
  candidates. Among them, scenarios of inelastic Dark Matter evade
  direct detection by suppressed elastic scattering off atomic nuclei. We
  introduce \emph{inelastic Dirac Dark Matter}, a new model with two
  Dirac fermions in the MeV-GeV mass range. At feeble couplings, dark matter can depart from chemical as well as kinetic equilibrium with
  the Standard Model before freeze-out. In this case, the freeze-out is driven by conversion processes like coscattering, rather than coannihilation. We show that inelastic
  Dirac relics are consistent with cosmological observations, in particular with nucleosynthesis and the cosmic microwave background. Searches
  for dark sectors at colliders and fixed-target experiments, in turn,
  are very sensitive probes. Compared to the strongly constrained
  pseudo-Dirac scenario, inelastic Dirac Dark Matter offers a new search
  target for existing and upcoming experiments like Belle II, ICARUS,
  LDMX and SeaQuest.}

\maketitle
\flushbottom

\clearpage

\section{Introduction}\label{sec:intro}
Direct detection has put pressure on thermal WIMPs. Current searches
for dark matter-nucleon scattering are extremely sensitive, and null
results have ruled out many scenarios of weakly interacting massive
particles (WIMPs) as candidates for cold thermal dark matter (DM)~\cite{Arcadi:2017kky,Roszkowski:2017nbc}.\footnote{Yet, the domain of
thermal dark matter is large and thermal WIMPs can still be viable
dark matter candidates~\cite{Leane:2018kjk,Lin:2019uvt,Coy:2021ann}.} An elegant option to
evade direct detection constraints is to consider \emph{inelastic Dark
  Matter}, where elastic nucleon scattering is absent or parametrically suppressed and inelastic up-scattering into a heavier dark partner is kinematically
suppressed~\cite{Tucker-Smith:2001myb}.

Inelastic dark matter has become a benchmark target for searches in particle physics and astrophysics~\cite{Beacham:2019nyx}. In a minimal realization, commonly dubbed iDM, the dark sector consists of a pseudo-Dirac state with two Majorana fields, $\xi_1$ and $\xi_2$, interacting with quarks and leptons via an abelian dark force by exchanging a dark photon $A'$~\cite{Izaguirre:2015zva}. This interaction drives the dark matter freeze-out in the early universe and the relic abundance is set through coannihilation $\xi_1\xi_2 \to A'^\ast \to f\bar{f}$ into Standard Model (SM) fermions $f$. The abundance measured today~\cite{Planck:2018vyg} favors iDM candidates in the MeV-GeV mass range. The phenomenology of this predictive scenario has been investigated in detail~\cite{Izaguirre:2015zva, Izaguirre:2015yja,Bramante:2016rdh,Izaguirre:2017bqb,Berlin:2018pwi,Berlin:2018bsc,Tsai:2019buq,Duerr:2019dmv,Duerr:2020muu,Kang:2021oes,Baryakhtar:2020rwy,Batell:2021ooj,CarrilloGonzalez:2021lxm,Bell:2021xff}, specifically for direct detection~\cite{Bramante:2016rdh,Baryakhtar:2020rwy,CarrilloGonzalez:2021lxm,Bell:2021xff}, at colliders~\cite{Duerr:2019dmv,Duerr:2020muu,Kang:2021oes}, and at fixed-target experiments~\cite{Izaguirre:2017bqb,Berlin:2018pwi,Berlin:2018bsc,Tsai:2019buq,Batell:2021ooj}. Taken together, searches for light dark particles in all three areas have excluded most of the parameter space of iDM. Inelastic dark matter from thermal freeze-out via coannihilation appears to be strongly constrained.

In this work, we introduce \emph{inelastic Dirac Dark Matter} (i2DM)
as a new model for feebly coupling dark matter. We promote the two
dark fermions in iDM to Dirac fields, one being charged and the other
one uncharged under a dark $U(1)$ gauge symmetry. The symmetry is
spontaneously broken by a Higgs-like mechanism, which causes the two
dark fermions to mix. As a result, the dark matter candidate $\chi_1$
interacts only feebly through the small mixing, while the coupling  of the dark partner
$\chi_2$ through the dark force is unsuppressed.

This moderate
modification of iDM leads to a very different cosmology: In i2DM, the
relic abundance can be set by partner annihilation $\chi_2 \chi_2 \to
f\bar{f}$ or coscattering $\chi_1 f\to \chi_2 f$, which is not an
option in iDM where thermal freeze-out is necessarily driven by
coannihilation. In contrast to the well-studied freeze-out through coannihilation and partner
  annihilation~\cite{Griest:1990kh,Edsjo:1997bg}, the role of conversion
  processes like coscattering has been investigated in specific dark matter
  scenarios only recently~\cite{Garny:2017rxs,DAgnolo:2017dbv,Garny:2018icg,DAgnolo:2018wcn,DAgnolo:2019zkf,Junius:2019dci,Herms:2021fql,Garny:2021qsr}. Our goal is to demonstrate
that i2DM is a new cosmologically viable candidate for feebly coupling inelastic dark matter in the
MeV-GeV range that can be probed at current and future experiments.

In Sec.~\ref{sec:model}, we introduce inelastic Dirac Dark Matter and explain the main characteristics of the model. The formalism is described in more detail in App.~\ref{app:model}. In Sec.~\ref{sec:freeze-out}, we analyze the density evolution of the dark sector in the early universe before freeze-out. We pay special attention to deviations from chemical and kinetic equilibrium, which occur for small couplings. In this regime, computing the relic abundance requires to solve a coupled system of Boltzmann equations and to keep track of the dark matter momentum distribution. Details on our calculations can be found in App.~\ref{app:CollisionTerm}. In
Sec.~\ref{sec:cosmo}, we investigate possible effects of i2DM on
astrophysical and cosmological observables. In particular, we discuss
the impact of the QCD phase transition, effects on the formation of
light elements, imprints on the cosmic microwave background, as well
as constraints from supernova cooling. The resulting bounds from
astrophysics and cosmology set a clear search target for i2DM. In
Sec.~\ref{sec:colliders}, we test this new dark matter target at
laboratory experiments. Signatures of i2DM strongly depend on the
lifetime of the dark partner. We find that fixed-target and flavor
experiments are most sensitive to dark fermions with long lifetimes
through searches for displaced decays, scattering, and missing
energy. In Sec.~\ref{sec:conclusions}, we conclude with an outlook to
future experiments that can conclusively test inelastic Dirac Dark
Matter.

\section{Inelastic Dirac Dark Matter}\label{sec:model}
We introduce a dark sector consisting of two Dirac fermions $\chi_1$ and $\chi_2$ with masses $m_1$ and $m_2$, interacting with the SM particles via a dark photon $A'$ with mass $m_{A'}$. The lighter of the dark fermions, $\chi_1$, serves as a dark matter candidate. The dark partner, $\chi_2$, will play a crucial role for the interactions between the dark and visible sectors.

Such a scenario can be constructed from a renormalizable theory with two fermion fields, both of them SM gauge singlets and one of them charged under a new abelian gauge symmetry $U(1)_D$. This symmetry is spontaneously broken by the vacuum expectation value of a dark scalar, which induces mixing between the dark fermions and gives the dark gauge boson a mass. The dark gauge boson kinetically mixes with the hypercharge field, thus acting as a mediator between the dark fermions and the SM fermions. For details on the model we refer the reader to App.~\ref{app:model}. A
 {\tt FeynRules} model for i2DM is available at \cite{FeynRulesModel}.

A priori, such a dark sector could be realized at any mass scale. Throughout this work we focus on dark particles in the MeV-GeV range, which could be resonantly produced at colliders and fixed-target experiments. The relevant interactions of the dark sector with visible matter are described by the Lagrangian
\begin{align}\label{eq:lagr}
	\mathcal{L} \supset \ e \epsilon A'_{\mu} \sum_f Q_f \bar{f} \gamma^\mu f  - g_D \Big(A'_{\mu} + \epsilon\,\frac{s_W}{c_W} Z_\mu\Big) \Big(\sin^2\theta  J^\mu_1 - \sin\theta \cos\theta J^\mu_{12} + \cos^2\theta J^\mu_2\Big),
\end{align}
where $f$ denotes the SM fermions with electric charge $Q_f$ in units
of the electromagnetic coupling $e$; $g_D$ is the coupling constant of the $U(1)_D$ symmetry; $s_W,c_W$ are the sine and cosine
of the weak mixing angle; $\theta$ is the mixing angle between the dark fermions; and $\epsilon$ parametrizes the kinetic mixing. The dark fermion currents are
\begin{equation}
J_1^\mu  = \bar{\chi}_1 \gamma^\mu \chi_1,\quad J_2^\mu = \bar{\chi}_2 \gamma^\mu \chi_2,\quad J^\mu_{12} = \bar{\chi}_1 \gamma^\mu \chi_2 + h.c.
\label{eq:J12}
\end{equation}
In Eq.~\eqref{eq:lagr}, we have kept only the leading terms in $\epsilon$. Modifications of the $Z$ boson couplings to SM fermions first occur at $\mathcal{O}(\epsilon^2)$, see Sec.~\ref{sec:ewpo}.

The mass mixing between the dark fermions, parametrized by $\theta$, determines the relative coupling strength of $\chi_1$ and $\chi_2$ to the dark photon. For $\theta \to 0$, the dark fermion $\chi_1$ decouples.
Throughout our analysis, we assume that the dark scalar that is responsible for the mixing is much heavier than the other dark particles and does not affect the observables we consider. In general, the presence of a light dark scalar could lead to interesting effects \cite{Duerr:2020muu,Baek:2020owl} and deserves a dedicated analysis.

The phenomenology of i2DM is described by six independent parameters
\begin{align}
    \big\{m_1,\Delta,m_{A'},\alpha_D,\epsilon,\theta\big\}
\end{align}
with $\alpha_D = g_D^2/(4\pi)$. The relative mass splitting between the dark fermions is defined as
\begin{equation}
\Delta = \frac{m_2 - m_1}{m_1}\,.
\label{eq:Delta}
\end{equation}
Throughout this work we focus on the mass hierarchy
\begin{align}\label{eq:mass-hierarchy}
    m_{A'} > 2 m_2\,,
\end{align}
so that decays of the dark photon into dark fermions are kinematically allowed.\footnote{For smaller $m_{A'}$, the phenomenology can change significantly. In particular, for $m_{A'} < m_1$, pair annihilations $\chi_1\chi_1 \to A'A'$ are important to set the relic abundance (see e.g.~\cite{DAgnolo:2015ujb}) and dark photons decay exclusively into SM fermions.} The total decay width of the dark photon is given by
\begin{align}
    \Gamma_{A'} & = \alpha_D \Gamma(A' \to \chi\overline{\chi}) + \epsilon^2 \alpha_e \Gamma(A' \to \rm{SM}),
\end{align}
where $\alpha_e=e^2/(4\pi)$ is the fine structure constant, while $\Gamma(A' \to \chi\overline{\chi})$ and $\Gamma(A' \to \rm{SM})$ denote the (normalized) decay rates into dark fermions and into leptons and hadrons, respectively. Dark photon decays into pairs of dark fermions, $\chi_i\overline{\chi}_j$, and leptons, $\ell_i \bar{\ell}_j$, are described by the kinematic function
\begin{align}
    \Gamma(m_i,m_j) & = \frac{m_{A'}}{2}
    \left[\left(1 - \frac{(m_i + m_j)^2}{m_{A'}^2}\right)\left(1 - \frac{(m_i - m_j)^2}{m_{A'}^2}\right)\right]^{\frac{1}{2}}\\\nonumber
    & \qquad\qquad \times \left(1 - \frac{(m_i - m_j)^2 - 4 m_i m_j}{2m_{A'}^2} - \frac{(m_i^2 - m_j^2)^2}{2 m_{A'}^4}\right).
\end{align}
Dark photon decays into hadrons can be computed by rescaling the leptonic decay rate with $e^+e^-$ data~\cite{Ilten:2018crw}. For $\epsilon^2\alpha_e \ll \alpha_D$, decays into SM particles are suppressed and the dark photon mostly decays into dark fermions. The corresponding decay rate is
\begin{align}
\Gamma(A' \to \chi\overline{\chi}) = \sin^4\theta\,\Gamma(m_1,m_1) + \sin^2(2\theta) \,\Gamma(m_1,m_2) + \cos^4\theta\, \Gamma(m_2,m_2)\,.
\end{align}
For $m_1 \approx m_2$, the branching ratios are determined to a good approximation by the dark fermion mixing, so that
\begin{align}\label{eq:brs}
    \mathcal{B}(A'\to \chi_1\overline{\chi}_1) \approx \sin^4\theta\,,\ \mathcal{B}(A'\to \chi_1\overline{\chi}_2,\chi_2\overline{\chi}_1) \approx \sin^2(2\theta)\,,\ \mathcal{B}(A'\to \chi_2\overline{\chi}_2) \approx \cos^4\theta\,.
\end{align}

The freeze-out dynamics relies on dark fermion annihilation into SM particles. Annihilations into leptons via $\chi_i \chi_j \to A'^\ast \to \ell^+\ell^-$ can be calculated in perturbation theory. Annihilations into hadrons can be predicted by re-scaling the cross section for annihilation into muons with the measured ratio~\cite{Ilten:2018crw,Zyla:2020zbs}
\begin{align}\label{eq:hadronic-decays}
    R(s) = \frac{\sigma(e^+e^- \to \text{hadrons})}{\sigma(e^+e^- \to \mu^+\mu^-)}\,.
\end{align}
The total cross section for dark fermion annihilation at a center-of-mass energy $\sqrt{s}$ is then given by
\begin{align}
    \sigma_{\chi_i \chi_j \to \text{SM}}(s) = \sum_{\ell = e,\mu,\tau} \sigma_{\chi_i \chi_j \to \ell^+\ell^-}(s) + \sigma_{\chi_i \chi_j \to \mu^+\mu^-}(s)\,R(s)\,.
\end{align}
In dark fermion interactions with SM fermions the dark photon acts as a virtual mediator. As a consequence, for $m_{A'} \gg m_{1,2}$ the scattering rates and decays of dark fermions scale as~\cite{Izaguirre:2015yja}
\begin{align}
  y = \epsilon^2\,\alpha_D \left(\frac{m_1}{m_{A'}}\right)^4.
\end{align}
As long as the dark photon is heavy compared to the momentum scale probed in observables, the dark sector interactions are described in terms of the four parameters
\begin{align}
    \{m_1,\Delta,y,\theta\}\,.
\end{align}
The phenomenology of i2DM crucially relies on the properties of the dark partner. For $m_2 \lesssim 1\,$GeV, the dark partner decays to almost 100\% into leptons~\cite{Duerr:2020muu}. The decay rate via a heavy virtual dark photon is given by
\begin{align}\label{eq:decay-rate}
   \Gamma(\chi_2 \to \chi_1 \ell^+ \ell^-) & = \frac{4 \alpha}{15 \pi} \tan^2\theta\cos^4\theta\, y\,m_1 \Delta^5 + \mathcal{O}\left(\frac{m_1^2}{m_{A'}^2}\right),
\end{align}
where we have neglected the lepton mass in the final state. We neglect hadronic decays in our analysis.

We will refer to this model as \emph{inelastic Dirac Dark Matter} (i2DM) to distinguish it from the widely studied scenario with Majorana fermions, often called pseudo-Dirac Dark Matter or simply inelastic Dark Matter (iDM)~\cite{Izaguirre:2015zva,Tsai:2019buq,Duerr:2019dmv,Duerr:2020muu,Kang:2021oes,Batell:2021ooj,CarrilloGonzalez:2021lxm}. iDM builds on a single pseudo-Dirac fermion charged under a new $U(1)_D$ force. The dark gauge symmetry is spontaneously broken by Majorana mass terms for the chiral components of the Dirac spinor. The resulting dark sector contains two Majorana fermions, $\xi_1$ and $\xi_2$, that couple to the dark photon mostly through inelastic interactions
\begin{align}\label{eq:idm}
\mathcal{L}_{\rm iDM} \supset - g_D A_\mu' J_{\rm iDM}^\mu\,,\qquad J_{\rm iDM}^\mu = i
 \big(\xi_2^\dagger \bar{\sigma}^\mu \xi_1 - \xi_1^\dagger \bar{\sigma}^\mu \xi_2\big)\,.
\end{align}

In both iDM and i2DM models, elastic scattering off atomic nuclei is suppressed, which strongly reduces the sensitivity of direct detection experiments. In i2DM, elastic dark matter interactions are additionally suppressed for small dark fermion mixing $\theta$. We will come back to direct detection in Sec.~\ref{sec:direct-detection}.

Despite similar predictions for nucleon scattering, i2DM and iDM feature very distinct dark matter dynamics in the early universe.
 The difference lies mostly in efficient $A'\chi_2\bar{\chi}_2$ interactions in i2DM, which are suppressed or completely absent in iDM. At first sight, this appears as a small modification. However, as we will show, the presence of partner interactions in i2DM has drastic effects on the dark matter freeze-out. We obtain new dark matter candidates with couplings that would be too small to explain the relic abundance from coannihilation, as in iDM.

\section{Freeze-out at feeble couplings}\label{sec:freeze-out}
Dark matter relics in the MeV-GeV range must be feebly coupled to the thermal bath in order to account for the observed DM abundance, $\Omega_\chi h^2 = 0.12$~\cite{Planck:2018vyg}. Moreover, viable scenarios of inelastic dark matter typically require a compressed spectrum of dark-sector particles. For i2DM, this leads to the parameter region of interest
\begin{equation}
  \{y, \theta, \Delta\} \ll 1\,.
  \label{eq:i2DMframework}
\end{equation}
Within this regime, the relic abundance can be set by various mechanisms. We consider scenarios where dark matter is in kinetic equilibrium with the thermal bath over a period of time before the relic abundance is set. The relic abundance should therefore be set by a freeze-out process, and we expect the dark fermions to be
non-relativistic around the freeze-out temperature\footnote{Other possible mechanisms for feebly coupled dark matter do exist. For instance, in the case of freeze-in~\cite{Hall:2009bx,Calibbi:2021fld}, dark matter was never in kinetic equilibrium and is usually
  relativistic at the time where its abundance freezes.}
\begin{align}
    T_{\rm fo} \ll 2 m_2 < m_{A'}.
\end{align}

In general, the freeze-out dynamics is determined by the evolution of the density distribution functions $f_i(x,q_i)$ for all relevant species $i$, expressed in terms of the dimensionless time and momentum variables
\begin{equation}
  x=\frac{m_1}{T}\quad{\rm and} \quad q_i=\frac{p_i}{T}\,.
  \label{eq:xq}
\end{equation}
Here $T$ is the temperature of the thermal bath, and we have chosen the
mass of the lightest dark fermion, $m_1$, to normalize $x$. The
norm of the three-momentum of species $i$ is denoted as $p_i =
|\vec{p}_i|$ and scales with the scale factor $a$ as $1/a$. The comoving number density, $Y_i$, is obtained from
the phase-space integration of the distribution function as
\begin{equation}\label{eq:comoving-density}
   Y_i(x) = \frac{n_i(x)}{s(x)}\,,\qquad n_i(x)= g_i \, \frac{m_1^3}{ x^{3}}\int \frac{d^3q} {(2 \pi)^3}\, f_i(x,q)\,,
\end{equation}
where $g_i$ is the number of degrees of freedom of species $i$. Both
the number density, $n_i(x)$, and the entropy density of the universe,
$s(x)$, scale as $1/a^3$. In what follows, we denote the number density of $\chi_i$ in kinetic equilibrium
and with zero chemical potential as
$n_i^{\rm eq}$. For later convenience, we define the ratios of number densities,
\begin{equation}
  r_i = \frac{n_i^{\rm eq}}{n^{\rm eq}}\,,
  \label{eq:ri-neq}
\end{equation}
where  $n^{\rm eq} = n_1^{\rm eq} + n_2^{\rm eq}$ is the total dark sector equilibrium number density.

The freeze-out temperature $T_{\rm fo}$ is determined by the time  $x_{\rm fo} = m_1/T_{\rm fo}$ at which the comoving dark matter density approaches the dark matter abundance $Y_0$ observed today,
 \begin{equation}
Y_1(x_{\rm fo}) = Y_0\,.
\label{eq:fo}
\end{equation}
In our numerical analysis, we determine $x_{\rm fo}$ by requiring that the dark matter density at freeze-out satisfies
 \begin{equation}\label{eq:freeze-out-temp}
x_{\rm fo} = \frac{m_1}{T_{\rm fo}}:\qquad  \frac{Y'_1(x_{\rm fo})}{Y_1^{\rm eq}(x_{\rm fo})} =  - 0.1\,.
 \end{equation}
 As we will see, i2DM dark matter candidates are not always in chemical and/or
kinetic equilibrium with the thermal bath around the freeze-out
temperature. As a result, freeze-out as defined in Eq.~\eqref{eq:fo}
does not necessarily coincide with chemical decoupling, as in the case
of vanilla WIMP dark matter~\cite{Bringmann:2006mu}. We therefore define the times $x_1$ and $x_2$, where $\chi_1$ and
  $\chi_2$ chemically decouple from the bath, corresponding to the decoupling temperatures $T_1$ and $T_2$.

 The time evolution of the density distribution function of a particle species $i$ is described by the Boltzmann equation
\begin{equation}
 E_i \, H x \, \partial_x f_i(x,q_i)={\cal C}[f_i(x,q_i),f_j(x,q_j)]\,.
  \label{eq:UnintBE}
\end{equation}
Here $E_i=E_i(x,q_i)$ is the energy associated with a species of mass
$m_i$ and momentum $q_i$, and $H=H(x)$ is the Hubble rate at time
$x$. In a radiation-dominated era, the Hubble rate scales as $H(x)\sim
x^{-2}$. The collision term ${\cal C}$ describes interactions of species $i$ with all other involved species $j$.

We emphasize that Eq.~(\ref{eq:UnintBE}) holds even if
species $i$ is not in kinetic and/or chemical equilibrium with the
thermal bath, but is only valid as long as the number of relativistic
degrees of freedom in the universe is constant. \footnote{For an i2DM
candidate within the parameter range defined in Eq.~\eqref{eq:parameters}, freeze-out happens between the QCD
phase transition and neutrino decoupling and Eq.~\eqref{eq:UnintBE}
applies.}

In what follows, we will refer to Eq.~\eqref{eq:UnintBE} as the
\emph{unintegrated Boltzmann equation}. By making several
simplifications, the set of $N$ partial integro-differential equations
for species $i,j = \{1,\dots N\}$ can be reduced to a set of ordinary
differential equations. See App.~\ref{app:CollisionTerm} for details.
 If all species are in
kinetic equilibrium the relic abundance can be calculated from the
time evolution of the number densities $n_i(x)$ from Eq.~\eqref{eq:comoving-density}~\cite{Griest:1990kh,Gondolo:1990dk}. The corresponding evolution equations are referred to as \emph{integrated Boltzmann equations}.

In the remainder of this section, we will discuss the freeze-out dynamics for i2DM. In Sec.~\ref{sec:relevant-processes}, we analyze all relevant annihilation, scattering and decay processes that
can play a role in setting the DM relic abundance. Depending on the relative importance of these processes, we encounter different phases of freeze-out. In Sec.~\ref{sec:phases}, we discuss these phases in detail and explain how to account for deviations from chemical and kinetic equilibrium when computing the relic abundance.

\subsection{Dark matter interactions}
\label{sec:relevant-processes}
For i2DM, the relevant interactions of the dark fermions with the thermal bath entering the collision term $\mathcal{C}$ in Eq.~\eqref{eq:UnintBE} are
\begin{align}
\text{(co)annihilation:} & \quad \chi_i\chi_j  \to A'^{\ast} \to f\bar{f}\,,\\\nonumber
\text{(co)scattering:} & \quad \chi_i f \to \chi_j f\,,\\\nonumber
\text{(inverse) decay:} & \quad \chi_i \to \chi_j f\bar{f}\,.
\end{align}
At temperatures below the GeV scale, $f$ denotes all (hadronized)
quarks and leptons that are in equilibrium with the thermal bath. To
determine the relevance of the various processes for DM freeze-out, we
investigate the thermally averaged interaction rates $\langle \Gamma
\rangle$ of the dark fermions with the bath. We define the interaction
rates for annihilation, scattering and decays of particle $i$ as
\begin{equation}\label{eq:interactions}
 \Gamth_{ij} = \frac{\gamma_{ij\to kl}}{n_i^{\rm eq}}, \qquad \Gamth_{i\to j}^{scat} = \frac{\gamma_{ik\to jl}}{n_i^{\rm eq}}, \qquad \Gamth_{i\to j}^{dec} = \frac{\gamma_{i\to jkl}}{n_i^{\rm eq}}\,.
\end{equation}
Here and below, $i,j = \{1,2\}$ denote the dark fermions $\chi_1$ and $\chi_2$, and $k,l$ label the SM fermions $f$. The brackets indicate the thermal average. The reaction densities $\gamma$ can be expressed as~\footnote{Reaction
  densities have been widely used in the context of leptogenesis, as in Ref.~\cite{Luty:1992un}. They allow us to define the relevant interaction rates in a thermal bath for the production of feebly coupled dark matter in a compact and unambiguous way~\cite{Frigerio:2011in,Junius:2019dci}.  For more details see App.~\ref{app:CollisionTerm}.}
\begin{eqnarray}
\gamma_{ab\to cd} = n_a^{\rm eq}n_b^{\rm eq}\,\langle \sigma_{ab\to cd}\,v_{ab}\rangle,\qquad
\gamma_{a\to bcd} = n_a^{\rm eq}\,\Gamma_{a\to bcd}\, \frac{K_1(m_a/T)}{K_2(m_a/T)} \,,\label{eq:gam-min}
\end{eqnarray}
where $a,b,c,d$ can denote particles from both the visible and the dark sector, $\langle\sigma_{ab\to cd}\,v_{ab}\rangle$ is
 the thermally averaged cross section with particles $a$ and $b$ in the initial state, $\Gamma_{a\to bcd}$ is the decay rate of particle $a$ in its rest
frame, and $K_1,\,K_2$ denote the modified Bessel functions of the first and second kind. The Moeller velocity is defined by
\begin{align}\label{eq:Moeller}
    v_{ab} = \sqrt{(p_a\cdot p_b)^2 - (m_a m_b)^2}/(E_a E_b)\,,
\end{align}
where $p_a,\,p_b$ denote the 4-momenta of particles $a$ and $b$, and $E_a,\,E_b$ are their energies in the center-of-mass frame.\footnote{In the text, by default $p_a$ refers to the norm of the 3-momentum. Only in the Moeller velocity or in a delta function that enforces 3-momentum {\it and} energy
  conservation, $p_a$ denotes the 4-momentum.}
 For simplicity, we label the reaction densities only by the involved dark fermions as
\begin{align}    \label{eq:gamij}
     \gamma_{ij} & \equiv \gamma_{ij \to kl} =\gamma_{kl\to ij}\,,\\\nonumber
     \gamma_{i\to j}^{scat} & \equiv \gamma_{ik\to jl} = \gamma_{jl\to ik}\,,\\\nonumber
     \gamma_{i\to j}^{dec} & \equiv \gamma_{i\to jkl} = \gamma_{jkl\to i}\,.
\end{align}
The relevance of the various interaction rates at a certain temperature can be inferred from their scaling with the time variable $x = m_1/T$ and with the model parameters. For non-relativistic dark sector particles $\chi_1$ and $\chi_2$ and relativistic involved SM fermions around freeze-out, the reaction
densities for (co)annihilations in thermal equilibrium scale as
\begin{align}\label{eq:annihilation}
 \gamma_{11} & = n_1^{\rm eq} n_1^{\rm eq}\, \langle\sigma(\chi_1 \chi_1 \to f\bar{f})\, v_{11} \rangle \propto x^{-3} e^{-2x}  \, y \, \sin^4\theta \\\nonumber
 \gamma_{12} & = n_1^{\rm eq} n_2^{\rm eq}\,\langle \sigma(\chi_1 \chi_2 \to f\bar{f})\, v_{12} \rangle \propto x^{-3}  e^{-2x}  \, y \,  \sin^2(2\theta)  \, e^{-x\Delta}
 \\\nonumber
  \gamma_{22} & = n_2^{\rm eq} n_2^{\rm eq}\, \langle\sigma(\chi_2 \chi_2 \to f\bar{f})\, v_{22} \rangle \propto x^{-3} e^{-2x} \, y \,\cos^4\theta \, e^{-2x\Delta}.
\end{align}
The (co)scattering reaction densities scale as
  \begin{align}\label{eq:scattering}
  \gamma_{1\to1}^{scat} & = n_1^{\rm eq} n_f^{\rm eq}\, \langle \sigma(\chi_1 f \to \chi_1 f)\, v_{1f} \rangle \propto x^{-9/2} e^{-x} \, y \,\sin^4\theta\,  v_{1f}^3\\\nonumber
    \gamma_{2\to1}^{scat} & = n_2^{\rm eq} n_f^{\rm eq}\,\langle\sigma(\chi_2 f \to \chi_1 f)\, v_{2f} \rangle \propto x^{-9/2} e^{-x} \, y \,\sin^2(2\theta) \, v_{2f}^3\, e^{-x \Delta}
    \\\nonumber
    \gamma_{2\to2}^{scat} & = n_2^{\rm eq} n_f^{\rm eq}\,\langle\sigma(\chi_2 f \to \chi_2 f)\, v_{2f} \rangle \propto x^{-9/2} e^{-x} \, y  \, \cos^4 \theta\, v_{2f}^3 \, e^{-x \Delta}.
  \end{align}
Three-body decays and inverse decays yield the reaction density
  \begin{equation}
    \gamma_{2\to1}^{dec} = n_2^{\rm eq}\,\Gamma(\chi_2 \to \chi_1 f\bar{f})\,\frac{K_1(x (1+\Delta))}{K_2(x (1+\Delta))} \propto x^{-3/2} e^{-x}\, y\,\sin^2(2\theta)\, \Delta^5 \,  e^{-x\Delta},
    \label{eq:decay}
  \end{equation}
given that $K_1(x)/K_2(x)\to 1$ for $x\gg 1$.\footnote{In Eqs.~\eqref{eq:annihilation}, \eqref{eq:scattering} and \eqref{eq:decay} we have neglected subleading contributions in $m_1^2/m_{A'}^2$, $m_f^2/m_{A'}^2$, $\Delta$ and $v_{ij}$, but include them in our numerical analysis.} For
processes with leptons in the final state we have calculated all rates analytically. To compute annihilations into hadrons, we rescale the cross section as described in Sec.~\ref{sec:model}, following Ref.~\cite{Duerr:2020muu}. For scattering processes we only include the dominant scatterings $\chi \ell^\pm \to
\chi \ell^\pm$ off leptons $\ell = \{e,\mu\}$, which are still relativistic at sub-GeV temperatures. We neglect scatterings off hadrons, whose number densities are Boltzmann-suppressed for masses above the muon mass.

In the expressions above, the variable $y$ determines the effective strength of the dark force and thereby the overall efficiency of dark sector interactions. As we will discuss in Secs.~\ref{sec:cosmo} and \ref{sec:colliders}, $y$ is constrained by cosmology and laboratory searches. In addition, the dark fermion mixing $\theta$ must be small to circumvent bounds from direct detection experiments. Therefore dark matter annihilation $\gamma_{11}$ and scattering $\gamma_{1\to 1}$ must be strongly suppressed. In the absence of further interactions, such a suppression would lead to an overabundance of dark matter today. As a consequence, feebly interacting i2DM candidates cannot be thermal WIMPs in the classical sense, where the relic abundance is determined by the WIMP pair-annihilation rate at freeze-out.

Indeed, for feebly coupling i2DM, dark matter pair
annihilation and scattering with the thermal bath play no role for the
temperature evolution of the dark matter density. Instead, the
evolution of $\chi_1$ is driven by interactions with the dark partner
$\chi_2$. The dark partner is kept in chemical and kinetic equilibrium with
the bath via efficient annihilation and scattering, driven by the
reaction densities $\gamma_{22}$ and $\gamma_{2\to 2}$. However, at
freeze-out all interaction rates of $\chi_2$ are exponentially Boltzmann-suppressed by powers of $e^{-x\Delta}$, compared to the interactions of $\chi_1$. For $\chi_2$ interactions to impact
the freeze-out of $\chi_1$, the mass difference $\Delta$ must be
small, as in Eq.~\eqref{eq:i2DMframework}. Viable scenarios of i2DM
with small dark matter couplings require a compressed spectrum of dark
fermions.

Conversions $\chi_1\leftrightarrow \chi_2$ play an essential role in
i2DM freeze-out; they keep dark matter in equilibrium with the thermal
bath and impact the evolution of the number density.  Both
coscatterings and decays contribute to the conversion rate of $\chi_i$,
\begin{align}\label{eq:conversion-rate}
   \Gamth_{i\to j} \equiv \Gamth_{i\to j}^{scat} + \Gamth_{i\to j}^{dec} = \frac{\gamma_{i\to j}^{scat} + \gamma_{i\to j}^{dec}}{n_i^{\rm eq}}\,.
\end{align}
Due to the respective scaling of the reaction densities with $x$, see
Eqs.~\eqref{eq:scattering} and \eqref{eq:decay}, the suppression of
the thermal rates at low temperatures is stronger for coscattering
($\gamma_{2\to 1}^{scat} \propto x^{-9/2}$) than for decays
($\gamma_{2\to 1}^{dec} \propto x^{-3/2}$). Depending on their
relative amplitude at a given temperature, either process can dominate
the conversion rate $\Gamth_{i\to j}$ and thus the evolution of
the number densities.

\begin{figure}
   \centering
  \includegraphics[scale=0.55]{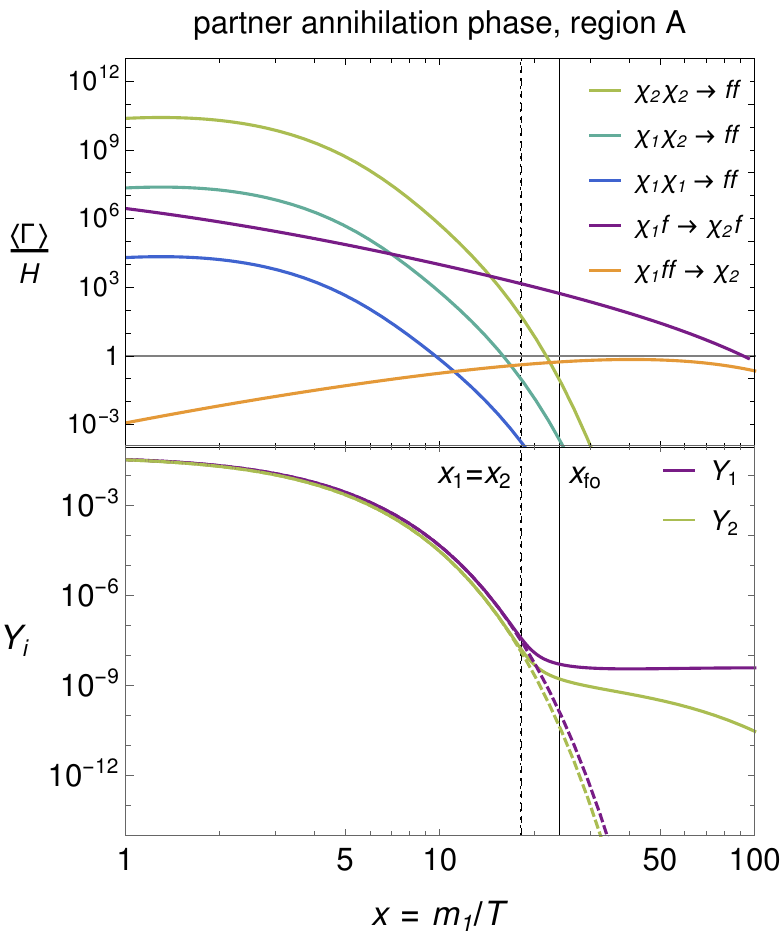}\hspace*{0.2cm}
  \includegraphics[scale=0.55]{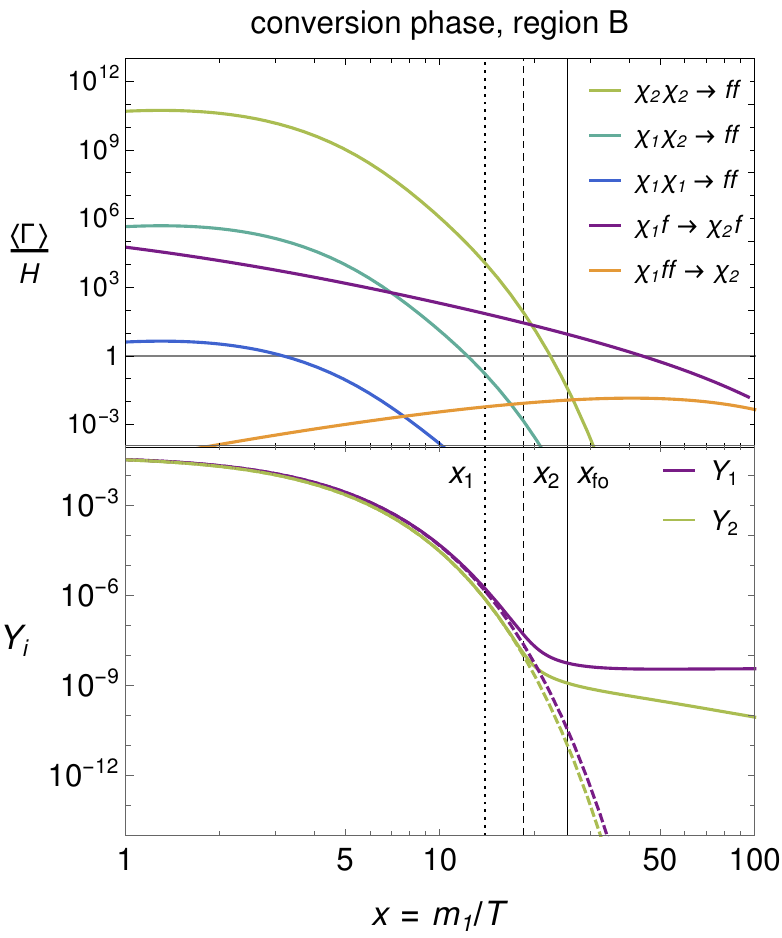}\vspace*{0.2cm}
  \includegraphics[scale=0.55]{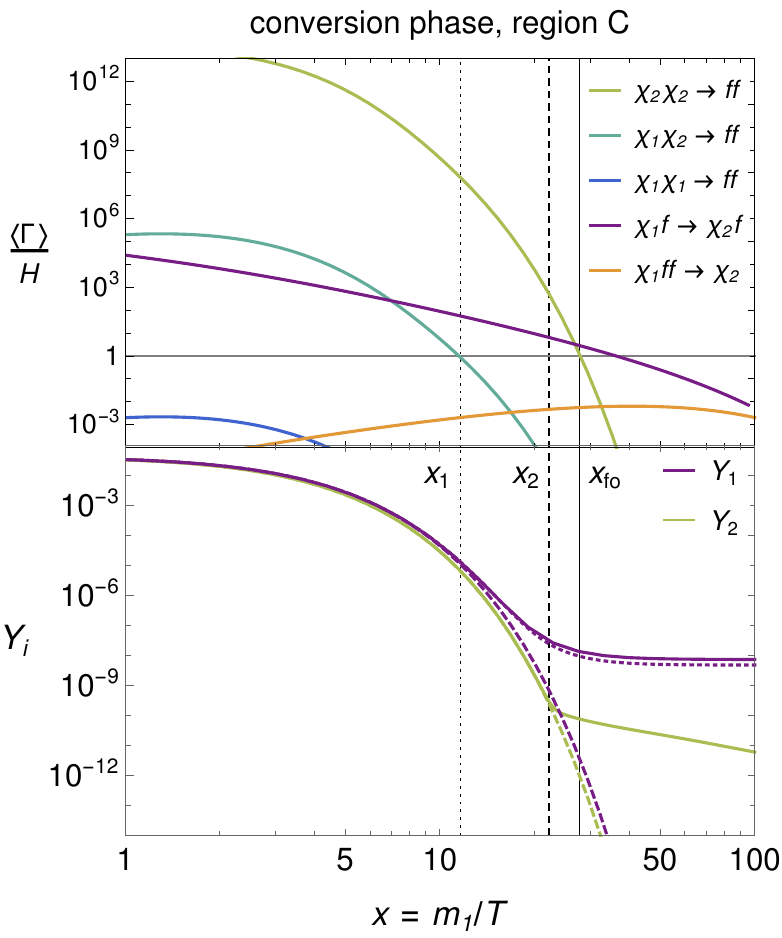}
  \caption{Evolution of the dark fermion interaction rates $\langle \Gamma \rangle$ and
    yields $Y$ as a function of $x=m_1/T$ for three i2DM benchmarks with
    $m_1 = 60\,$MeV, $\Delta = 0.05$ and $\alpha_D = 1/4\pi$. The
    relic abundance is set by partner annihilation (upper left plot),
    conversions in kinetic equilibrium (upper right plot) and
    conversions out of kinetic equilibrium (lower plot), happening in
    regions (A), (B) and (C), respectively.  The model
    parameters $(\tan\theta, y)$ are set to $(0.02,4\cdot
    10^{-10})$, $(0.003, 10^{-9})$ and $(10^{-4},4\cdot 10^{-7})$. Top panels: interaction rates $\langle \Gamma\rangle / H \equiv \{\langle
    \Gamma\rangle_{ij},\langle \Gamma\rangle_{i\to j}\}\cdot n_i^{\rm eq}/(n_1^{\rm eq} H)$, $i,j=\{1,2\}$, normalized to the $\chi_1$ equilibrium
    number density and the Hubble rate. Bottom panels: comoving number densities $Y_i$ of dark fermions $\chi_i$ (solid purple and green curves) and equilibrium number densities $Y_i^{\rm eq}$ (dashed curves). In the lower plot, the naive $\chi_1$ number density obtained from integrated Boltzmann equations (dotted purple
    curve) is shown for comparison with the correct result from unintegrated Boltzmann equations (solid curve). See Sec.~\ref{sec:conversion} for details. Vertical black lines indicate the times for
    $\chi_1$ chemical decoupling (dotted), $\chi_2$ chemical decoupling
     (dashed), and $\chi_1$ freeze-out (solid).}
  \label{fig:ratiogamH}
 \end{figure}

To illustrate the impact of the various processes on i2DM, in Fig.~\ref{fig:ratiogamH} we show the time
evolution of the dark sector interaction rates $\langle \Gamma \rangle$ and the yield $Y$ for three
i2DM benchmarks. The three benchmarks belong to different phases of freeze-out, which we will discuss in detail in Sec.~\ref{sec:phases}.

\subsection{Phases of freeze-out}
\label{sec:phases}
From the discussion in Sec.~\ref{sec:relevant-processes}, it becomes clear that
the freeze-out dynamics should be very sensitive to the parameters $y,\,
\theta$ and $\Delta$. When successively decreasing the dark interaction strength $y$, we identify three different \emph{phases of freeze-out}, distinguished by the processes that set the dark matter relic
abundance:
\begin{enumerate}
\item\label{item:coan} {\bf coannihilation phase}: $\Omega_\chi h^2$ set by $ \chi_1 \chi_2 \leftrightarrow f\bar{f}$ and $\chi_2 \chi_2 \leftrightarrow f\bar{f}$,
\item\label{item:partn} {\bf partner annihilation phase}: $\Omega_\chi h^2$ set by $ \chi_2 \chi_2 \leftrightarrow f\bar{f}$,
\item\label{item:conv} {\bf conversion phase}: $\Omega_\chi h^2$ set by $\chi_1 f \leftrightarrow \chi_2 f$ and/or $\chi_2 \leftrightarrow \chi_1 f\bar{f}$.
\end{enumerate}
In addition,  the thermal history of the dark matter candidate
 depends on whether departures from chemical or kinetic
equilibrium with the bath have occurred prior to freeze-out. This
critically depends on the efficiency of $\chi_1\leftrightarrow \chi_2$ conversions at the time $x_2$ at which $\chi_2$ chemically decouples. We distinguish between three regions of parameter space,
 where the following conditions are satisfied:
  \begin{eqnarray}
        (\text{A}) & &\qquad\  \ \frac{\Gamth_{1\to 2}}{H}(x_{2}) \gtrsim 100
        \label{eq:equ}\\
        (\text{B}) & &\ 10 \lesssim \frac{\Gamth_{1\to 2}}{H}(x_{2}) \lesssim 100 \label{eq:chemdec}\\
        (\text{C}) & &\qquad\ \ \frac{\Gamth_{1\to 2}}{H}(x_{2})  \lesssim  10 \,,\label{eq:kindec}
    \end{eqnarray}
where $\Gamth_{1\to 2}$ denotes the conversion rates from Eq.~\eqref{eq:conversion-rate} and $H(x_2)$ is the Hubble expansion at the time $x_2$. These three regions allow us to systematically study the effects of dark matter chemical and kinetic decoupling before freeze-out on the relic abundance.

Chemical decoupling of $\chi_2$ occurs when its
      (co)annihilation rate drops below the Hubble rate. If
      conversions $\chi_1 \leftrightarrow \chi_2$ are efficient around
      $x_2$, the decoupling time can roughly be estimated
        using
         \begin{align}
      r_2\big(\Gamth_{22}(x_2) + 2\Gamth_{21}(x_2)\big) \approx  H(x_2)\,,
          \label{eq:x22dec}
         \end{align}
 and $\chi_2$ decouples at the same time as $\chi_1$. However, if conversions are absent, $\chi_2$
  chemically decouples around
 \begin{align}
  \Gamth_{22}(x_2) \approx H(x_2)\,.
 \end{align}
Numerically we determine the time $x_i$ where $\chi_i$ chemically decouples by requiring that the density yield deviates from equilibrium by 20\%,
\begin{align}\label{eq:chi-dec}
 x_i = \frac{m_1}{T_i}:\qquad \frac{Y_i(x_i)}{Y_i^{\rm eq}(x_i)} = 1.2\,,\qquad i=\{1,2\}.
 \end{align}

\begin{figure}
  \centering
\includegraphics[scale=0.65]{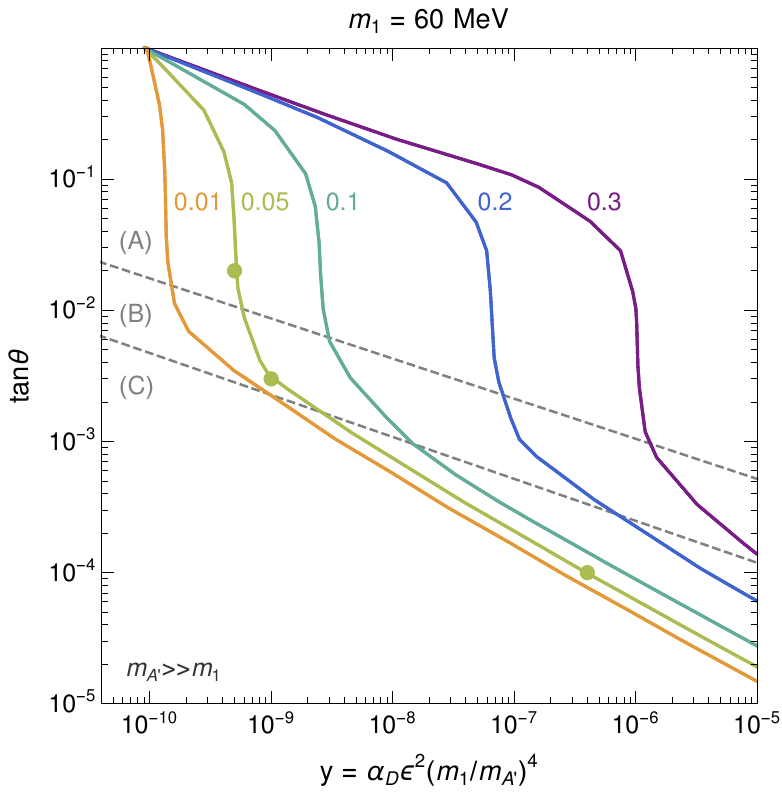}\vspace*{0.5cm}
\includegraphics[scale=0.65]{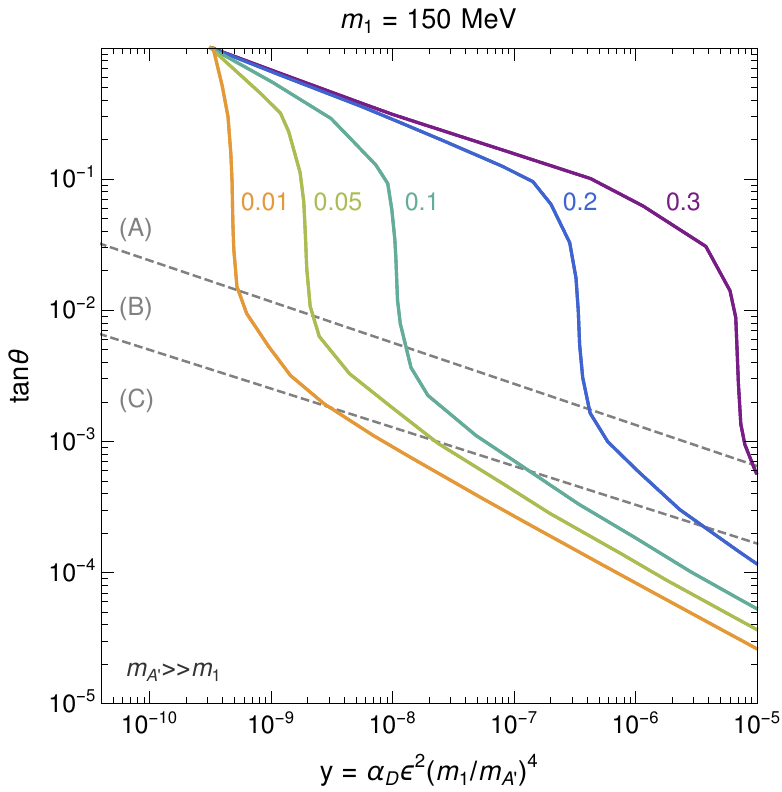}
\caption{Dark matter relic abundance for i2DM as a function of the dark interaction strength, $y$, and
  the dark fermion mixing, $\tan\theta$. The dark matter mass is fixed to $m_1 = 60\,$MeV (top) and 150 MeV (bottom). The observed abundance $\Omega_{\chi} h^2=0.12$ is obtained along the colored contours for different dark fermion mass
  splittings $\Delta=0.01 \dots 0.3$. The decoupling regions (A), (B) and (C) are separated by dashed gray lines, which satisfy Eqs.~(\ref{eq:kindec1}) and~(\ref{eq:kindec2}). In region (A), the dark sector is in chemical and kinetic equilibrium with the bath. In regions (B) and (C), $\chi_1$ successively decouples from chemical and kinetic
  equilibrium with $\chi_2$ prior to freeze-out. In the top panel, the
  bullets on the green contour for $\Delta=0.05$ mark the position of the
  three i2DM benchmarks displayed in Fig.~\ref{fig:ratiogamH}.
    \label{fig:phases}}
\end{figure}

The classification made above allows us to understand the density evolution of the dark fermions shown in Fig.~\ref{fig:ratiogamH}. The three i2DM benchmarks correspond to the freeze-out phases of partner annihilation (upper left plot) and conversion (upper right and lower plots). Partner
 annihilation is relevant in region (A), while conversion
 can prevail either in region (B) in kinetic equilibrium (upper right
 plot), or in region (C) beyond kinetic equilibrium (lower plot). The interplay between the different freeze-out phases and decoupling regions is shown in Fig.~\ref{fig:phases} as a function of the model parameters $y$ and $\tan\theta$ for two scenarios with fixed dark matter masses.
  In the upper plot of Fig.~\ref{fig:phases}, the three benchmark i2DM scenarios from
 Fig.~\ref{fig:ratiogamH} are marked as green bullets. Below we first discuss Fig.~\ref{fig:ratiogamH} in detail
 and then turn to Fig.~\ref{fig:phases}.

 In Fig.~\ref{fig:ratiogamH}, all benchmarks correspond to fixed parameters $m_1 = 60\,$MeV, $\Delta = 0.05$ and $\alpha_D = 1/4\pi$. In each of the plots, the top panel shows the evolution of the various interaction rates $\langle\Gamma\rangle$, normalized to the Hubble rate. The conversion rate $\Gamma_{1\to 2}$ that distinguishes regions (A), (B), and (C) is driven by $y\tan^2 \theta$, see Eq.~\eqref{eq:scattering}. It decreases when going
 from partner annihilation in region (A) (upper left plot) to
 conversion beyond kinetic equilibrium in region (C)
 (lower plot). The relative scaling of the interaction rates is determined by the
 reaction densities from Eqs.~\eqref{eq:annihilation},
 \eqref{eq:scattering} and \eqref{eq:decay} discussed in
 Sec.~\ref{sec:model}. In particular, the reaction densities for
 (co)annihilations, $\gamma_{ij}\sim e^{-2x}$, drop faster at low
 temperatures than scattering and decays, $\gamma_{i\to j}\sim
 e^{-x}$. The relative strength of (co)annihilations depends
 exponentially on the mass splitting $\Delta$ and also on the dark
 fermion mixing $\theta$. Due to the small splitting and mixing in the three benchmarks,
 partner annihilation dominates (light green), followed by coannihilation (dark green) and
 suppressed dark matter annihilations (blue). As mentioned in
 Sec.~\ref{sec:model}, conversions through scattering
 (purple) decrease faster with time than inverse decays (orange). Around freeze-out, however, conversions dominate over decays in all three benchmarks.

In the bottom panels of Fig.~\ref{fig:ratiogamH}, we show the time evolution of the dark
fermion comoving number densities, $Y_i(x)$, (solid) and the equilibrium yield, $Y_i^{\rm  eq}(x)$, for comparison (dashed). In the lower plot, we also indicate the evolution that is obtained when neglecting the kinetic decoupling of dark matter (dotted). To guide the eye, we highlight the times for freeze-out and chemical decoupling with black vertical lines, determined by Eqs.~\eqref{eq:freeze-out-temp} and~\eqref{eq:chi-dec}.

 We now turn our attention to Fig.~\ref{fig:phases}, which illustrates
 the different phases of freeze-out for i2DM as a function of the dark
 interaction strength $y$ and the dark fermion mixing $\tan\theta$ for
 two fixed dark matter masses $m_1=60$ MeV (top) and 150 MeV
 (bottom). The regions (A), (B) and (C), corresponding to decreasingly
 efficient conversions as suggested  by
 Eqs.~\eqref{eq:equ}-\eqref{eq:kindec}, are delineated with dashed
 gray lines. The exact relations between the conversion rate and
   the Hubble rate along these lines are given in Eqs.~\eqref{eq:kindec1} and~\eqref{eq:kindec2}. The observed relic
 abundance $\Omega_{\chi} h^2 = 0.12$ is obtained along the solid
 colored contours in the $(y,\tan \theta)$ plane for fixed values of
 the mass splitting $\Delta$. When increasing the dark matter mass
 $m_1$, the contours shift to the right, meaning that the observed
 abundance is obtained for larger values of $y$. This is easily
 understood, as all (co)annihilation and conversion rates scale as
 $\Gamth_{ij}, \Gamth_{i\to j}\propto y/m_1^2$. In the upper part of
 the plots, all contours converge and the mass splitting $\Delta$
 plays no role in setting the relic abundance. Here the abundance is
 set by pair annihilations $\chi_1 \chi_1 \to f\bar{f}$, see
 Eq.~(\ref{eq:annihilation}). In Sec.~\ref{sec:colliders}, we will see
 that laboratory searches exclude this region of parameter space.  As
 a result, we focus on dark matter candidates with small couplings,
 corresponding to the phases of coannihilation and partner
 annihilation in region (A), and on conversions in regions (B) and
 (C).

In region (A), efficient conversion rates satisfying Eq.~(\ref{eq:equ})
keep $\chi_1$ in chemical and kinetic equilibrium with the thermal
bath until freeze-out. In particular, efficient conversions ensure
that $\chi_1$ and $\chi_2$ have equal chemical potentials, so that
their number densities are related by
   \begin{align}\label{eq:chemical-eq}
      \frac{n_1}{n_2} = \frac{n_1^{\rm eq}}{n_2^{\rm eq}}\,.
  \end{align}
 In this case, the freeze-out conditions are similar to the ones of a
 thermal WIMP and the observed relic dark matter abundance is obtained
 for a freeze-out time~\cite{Griest:1990kh,Gondolo:1990dk,
   Edsjo:1997bg}
\begin{equation}
 x_{\rm WIMP}\simeq 25\,.
\label{eq:xfoWIMP}
\end{equation}
This is illustrated by the density evolution for the benchmark in the
upper left plot of Fig.~\ref{fig:ratiogamH}, which corresponds to the
upper green bullet in the upper plot of Fig.~\ref{fig:phases}. At large $\tan \theta$ of region (A), we note the rapid bending of the curves. This happens when coannihilations and pair annihilations of the lightest dark state start playing significant role at decoupling of $\chi_1$. The relative impact of the mentioned reactions is not important for our discussion as long as condition (A) is satisfied.

When entering in region (B), defined by Eq.~\eqref{eq:chemdec}, conversion processes are less efficient and the rates
$\Gamth_{1\to 2}(x_2)$ are about ten to one hundred times larger than the Hubble rate. As a result, the dark matter density departs from
chemical equilibrium prior to freeze-out. The effect is visible in the second benchmark displayed in the
upper right plot of Fig.~\ref{fig:ratiogamH}, corresponding to the
second green bullet in the upper plot of Fig.~\ref{fig:phases}.

Once the conversion rate $\Gamth_{1\to 2}(x_2)$ is further suppressed, dark matter cannot be
expected to be kept in kinetic equilibrium until freeze-out. We
enter region (C), defined by Eq.~\eqref{eq:kindec} and illustrated by a benchmark in the lower plot
of Fig.~\ref{fig:ratiogamH}, corresponding to the lower green bullet
of the upper plot of Fig.~\ref{fig:phases}. To quantify the effect of kinetic decoupling, in
Fig.~\ref{fig:ratiogamH} we display the physical dark
matter yield obtained by respecting kinetic
decoupling (solid purple curve) compared to the same yield obtained when neglecting kinetic decoupling (dotted purple curve).

In what follows, we will discuss the physics of the three freeze-out phases in detail, paying particular attention to non-equilibrium effects in the
conversion phase. Further details about the computation of the dark
matter relic abundance in the presence of chemical and kinetic
decoupling can also be found in App.~\ref{app:CollisionTerm}.

\subsubsection{Coannihilation}
\label{sec:coannihilation}
Coannihilation sets the relic abundance when the rate of $\chi_1\chi_2
\leftrightarrow f \bar f$ is larger than the Hubble rate and dominates over dark matter pair annihilations $\chi_1\chi_1 \leftrightarrow f \bar f$ around the freeze-out time, i.e., when
  \begin{align}
    \Gamth_{12}(x_{\rm fo}) > \Gamth_{11}(x_{\rm fo})\,.
\label{eq:Gamcoa}
  \end{align}
  According to Eqs.~\eqref{eq:interactions} and \eqref{eq:annihilation}, in i2DM the ratio of these rates scales as
  \begin{align}\label{eq:relative-annihilation}
 \frac{\Gamth_{11}(x)}{\Gamth_{12}(x)} \sim \frac{n_1^{\rm eq}}{n_2^{\rm eq}}\, \tan^2\theta \sim e^{x\Delta}
   \tan^2\theta\,.
  \end{align}
   One might deduce that coannihilation sets the relic abundance if $e^{x_{\rm fo}\Delta}\tan^2\theta
   \lesssim 1$. However, this is only a necessary condition, because in this regime the relic abundance could also be driven by partner annihilation $\chi_2\chi_2 \to f\bar{f}$. To determine the relative impact of partner annihilation, one needs to consider the \emph{weighted} ratio of coannihilation and
   partner annihilation rates, $r_1\Gamth_{12}(x)/r_2\Gamth_{22}(x)$.\footnote{The effective
     coannihilation and partner annihilation rates entering in the
     computation of the dark matter relic abundance have to be
     weighted by $r_i$, the ratio of the dark species equilibrium
     densities to the total one, when the dark sector species are in
     chemical equilibrium, see e.g.~\cite{Griest:1990kh}.} In i2DM,
   this ratio scales as $e^{x\Delta}\tan^2\theta$, exactly as in Eq.~(\ref{eq:relative-annihilation}). As a result, in the
   coannihilation phase the comoving number density $Y_1(x)$
   freezes once the weighted sum of coannihilation and partner
   annihilation drops below the Hubble rate,\footnote{This relation assumes that $\chi_1$ and $\chi_2$ are both in kinetic and chemical equilibrium prior to freeze-out.}
  \begin{align}\label{eq:fo-coa}
     2r_1 \Gamth_{12}(x_{\rm fo}) + r_2 \Gamth_{22}(x_{\rm fo}) \approx H(x_{\rm fo})\,,\qquad x_{\rm fo}\simeq x_{\rm WIMP}\,.
  \end{align}
The relic dark matter abundance is set around the freeze-out
temperature of a thermal WIMP. Freeze-out through coannihilation is realized in the upper
part of region (A) in Fig.~\ref{fig:phases}. Due
to the scaling $\Gamth_{12} \sim y\,\sin^2(2\theta)$, the relic
abundance contours tend to larger interaction strength $y$ at smaller
mixing $\theta$ in this regime.

In the coannihilation phase, i2DM resembles iDM, where the $\xi_1 - \xi_2$ couplings of Eq.~\eqref{eq:idm} prevail and coannihilations $\xi_1\xi_2 \to f\bar{f}$ are the dominant number-changing interactions of the dark matter candidate $\xi_1$ with the bath. However, lowering the dark interaction strength $y$ leads to inefficient coannihilation around freeze-out. In iDM, this results in an overabundance of dark matter. In i2DM, suppressed coannihilations can be compensated by efficient partner annihilations and conversions, thus explaining the observed dark matter abundance even if the dark sector is feebly coupled.

\subsubsection{Partner annihilation}
\label{sec:partner-annihilation}
If the coannihilation rate is suppressed compared to the Hubble rate around $x=x_{\rm WIMP}$,
the relic DM abundance can be set by partner annihilations $\chi_2\chi_2 \to f\bar{f}$. In this phase, the freeze-out condition reads\footnote{This condition assumes that Eq.~(\ref{eq:equ}) holds.}
 \begin{align}
r_2\,\Gamth_{22}(x_{\rm fo}) \approx H(x_{\rm fo})\,, \qquad  x_{\rm fo}\simeq x_{\rm WIMP}\,.
\label{eq:Gampart}
 \end{align}
As in the coannihilation phase, the freeze-out time is fixed to $x_{\rm fo} \simeq x_{\rm WIMP}$, up to a moderate logarithmic dependence on the
model parameters~\cite{Griest:1990kh}. In Fig.~\ref{fig:phases}, the
phase of partner annihilation lies in region (A) and is characterized
by vertical lines. Due to the scaling $r_2\langle \Gamma_{22} \rangle
\sim y \cos^4\theta\, e^{- 2 x\Delta}$, the relic dark matter
abundance is essentially independent of the mixing as $\cos^4\theta
\approx 1$. To obtain the observed abundance, variations
of the dark interaction strength $y$ can be compensated by the mass
splitting $\Delta$, as illustrated by the
various contours. In the partner annihilation phase,
$\chi_1\leftrightarrow \chi_2$ conversions have to be efficient enough to satisfy the decoupling condition from Eq.~(\ref{eq:equ}). As a consequence, $\chi_1$ and
$\chi_2$ chemically decouple from the bath around the same time.  This
is visible in the upper left plot of Fig.~\ref{fig:ratiogamH}. Notice that $\chi_2 \to
\chi_1$ decays happen on a time scale shorter than the period of
chemical decoupling and do not affect the freeze-out time $x_{\rm fo} \approx x_{\rm WIMP}$.

Whenever coannihilation or partner annihilation set the
relic abundance, dark matter is in chemical and kinetic
equilibrium until decoupling. In this case, the evolution equations
from Eq.~\eqref{eq:UnintBE} can be reduced to one single
integrated Boltzmann equation, written in terms of the total number density of all dark particles, $n=\sum_i n_i$, as commonly used for
thermal WIMPs~\cite{Griest:1990kh,Edsjo:1997bg}. The dark matter
abundance can be computed with any of the available Boltzmann
solvers~\cite{Belanger:2018ccd,Bringmann:2018lay,Ambrogi:2018jqj},
which explicitly make use of Eq.~\eqref{eq:chemical-eq}. For the
computations in this work we have used our own Boltzmann solver. We have
verified that our results for the relic abundance in region (A) in
Fig.~\ref{fig:phases} and in the left panel of
Fig.~\ref{fig:ratiogamH} agree with the results obtained from
micrOMEGAs~\cite{Belanger:2018ccd}.

  \subsubsection{Conversion}
  \label{sec:conversion}
As we discussed above, partner annihilation can set the relic dark matter
abundance even if $\chi_1\chi_1\to f\bar{f}$ and $\chi_1 \chi_2\to f\bar{f}$ annihilations are
suppressed. If the dark fermion mixing $\theta$ is very small,
$\chi_1\leftrightarrow \chi_2$ conversion rates can be comparable to
or even fall below the Hubble rate around $\chi_2$ chemical
decoupling, as in Eqs.~\eqref{eq:chemdec} and \eqref{eq:kindec}. The dark matter abundance is now set by
conversion processes, i.e., by coscattering and/or (inverse)
decays.

In i2DM, conversion-driven freeze-out can occur in regions (B) and (C) in Fig.~\ref{fig:phases}, Here coscattering dominates over decays in the thermal history around freeze-out. Due to the scaling $\Gamth_{1\to 2} \propto \sin^2(2\theta)$,
the contours of constant $\Omega_\chi h^2$ are sensitive to the mixing
angle $\theta$.  For a fixed interaction
strength $y$ and mass splitting $\Delta$, the relic abundance in this regime is generally larger than what one would expected from
partner annihilation. The reason is that $\chi_1\leftrightarrow
  \chi_2$ conversions are less efficient and the dark matter yield
$Y_1(x)$ can start to deviate from the equilibrium yield
$Y_1^{\rm eq}(x)$ well before the freeze-out time $x=x_{\rm
    fo}$ ~\cite{Garny:2017rxs,DAgnolo:2018wcn}. The latter effect is
illustrated in the upper right and lower plots of
Fig.~\ref{fig:ratiogamH}. The increased yield has to be compensated by a larger
interaction strength $y$, causing the contours in
Fig.~\ref{fig:phases} to bend towards the lower right corner. Notice
that efficient partner annihilation is essential for conversions to
explain the observed relic abundance. Suppressed partner annihilations
would result in an overabundance of dark matter.

To describe deviations of the dark matter density from chemical and kinetic equilibrium within the conversion phase, it is convenient to distinguish
three key moments:
\begin{enumerate}
\item the time $x_1$ at which $\chi_1$ chemically decouples from the bath;
\item the time $x_{2}$ at which $\chi_2$ chemically decouples from the bath;
\item the dark matter freeze-out time $x_{\rm fo}$, which can differ from $x_{\rm WIMP} \simeq 25$ in the conversion phase.
\end{enumerate}
Numerically, we determine $x_1$, $x_2$ and $x_{\rm fo}$ using Eqs.~\eqref{eq:chi-dec} and~\eqref{eq:freeze-out-temp}. Deviations of the dark matter density from chemical equilibrium around freeze-out typically occur for
\begin{align}
x_1 \lesssim x_2 < x_{\rm fo}\,.
\end{align}
Deviations from chemical \emph{and} kinetic equilibrium can occur for
\begin{align}
x_1 \ll x_2 < x_{\rm fo}\,.
\end{align}
Thanks to efficient elastic scattering, the dark partner $\chi_2$ is kept in kinetic equilibrium with the bath throughout the dark matter freeze-out process and in particular for $x > x_2$ after $\chi_2$ chemical decoupling.

\paragraph{Deviations from chemical equilibrium} If $x_1 \lesssim x_2$,
 conversions are barely efficient during $\chi_2$ chemical decoupling,
 as in Eq.~\eqref{eq:chemdec}. This scenario corresponds to region (B)
 in Fig.~\ref{fig:phases}. Numerically we find that the boundary between regions (A) and
   (B) corresponds to
  \begin{align}
\Gamth_{1\to 2}(x_2)= 200\, H(x_2)\,,
\label{eq:kindec1}
  \end{align}
  using Eq.~(\ref{eq:chi-dec}) to evaluate $x_2$. Below this
  boundary, the $\chi_1$ density departs from chemical equilibrium
  before $\chi_2$. This effect is illustrated for a benchmark in the
  upper right plot of Fig.~\ref{fig:ratiogamH}, where the freeze-out
  process for $Y_1(x)$ terminates around $x_{\rm fo} > x_1$. At that
  time, we expect that conversions are still sufficiently active to
  keep $\chi_1$ in kinetic equilibrium with the thermal bath via
  $\chi_2$. In particular, we assume that the dark matter distribution
  function $f_1(x,q)$ is well approximated by the Boltzmann
  distribution $f_1^{\rm eq}(x,q)$ throughout the entire evolution
  process.\footnote{Conversion-driven freeze-out with deviations from
    chemical equilibrium was studied before in
    Ref.~\cite{Garny:2017rxs}, where $f_1(x,q)$ was observed to depart
    from $f_1^{\rm eq}(x,q)$ prior to dark matter freeze-out. In i2DM,
    however, we expect that the distribution
    function of $\chi_1$ resembles $f_1^{\rm eq}(x,q)$ before
    freeze-out, because $\chi_1$ is kept in kinetic equilibrium at
    early times via $\chi_2 \to \chi_1$ decays and coscatterings (with
    $\chi_2$ being in kinetic equilibrium with the bath), see
    Fig.~\ref{fig:ratiogamH}. This was not the case for the dark
    matter model studied in Ref.~\cite{Garny:2017rxs}.}

 To account for deviations from chemical equilibrium in region (B),
the Boltzmann equations commonly used for (co)annihilating
 dark matter~\cite{Griest:1990kh} have to be supplemented by
 explicitly including the conversion rate in the coupled system of
 evolution equations for $Y_{1}(x)$ and $Y_2(x)$. For details on our implementation of the
 Boltzmann equations we refer the reader to App.~\ref{sec:chem-dec}.

 \paragraph{Deviations from chemical and kinetic equilibrium}  For $x_1 \ll x_{2}$, conversions become inefficient before $\chi_2$ chemically decouples, and the condition of Eq.~(\ref{eq:kindec}) is satisfied. This scenario corresponds to region (C) in
  Fig.~\ref{fig:phases}. Numerically, we find that the boundary between regions (B) and (C) is given by
    \begin{align}
\Gamth_{1\to 2}(x_2)= 20\, H(x_2)\,,
\label{eq:kindec2}
\end{align}
with $x_2$ evaluated with Eq.~\eqref{eq:chi-dec}. Below this line, the mixing $\theta$ is so small that conversions fail to keep dark matter in kinetic equilibrium. This effect is visualized in the lower panel of Fig.~\ref{fig:ratiogamH}: The dark matter yield $Y_1(x_{\rm fo})$ including departures from kinetic equilibrium (solid curve) is larger than under the assumption of kinetic equilibrium (dotted curve). The time between dark matter chemical decoupling and freeze-out is now stretched over a larger range between $x_1$ and $x_{\rm fo}$.

In region (C), the phase-space density $f_1(x,q)$ is expected to
deviate significantly from a Maxwell-Boltzmann distribution for $x >
x_1$. To account for deviations from kinetic equilibrium, the
unintegrated Boltzmann equations from Eq.~\eqref{eq:UnintBE} have to
be solved for the momentum-dependent phase-space density of
$\chi_1$~\cite{DAgnolo:2017dbv,Garny:2017rxs}. Further details are
given in App.~\ref{sec:kin-dec}. Using our own implementation, we
obtain the solid purple curve in the lower plot of
Fig.~\ref{fig:ratiogamH}.  For comparison, we show the
evolution of $Y_1(x)$ obtained using the integrated Boltzmann
equations relevant in region (B), but neglecting deviations from kinetic
equilibrium (dotted line).  The deviations are modest, ranging around 35\%.\footnote{We have
 checked explicitly that using unintegrated Boltzmann equations of
  App.~\ref{sec:kin-dec} and the integrated equations of
  App.~\ref{sec:chem-dec} the resulting contours are
     very similar to the ones in Fig.~\ref{fig:phases}.}

Throughout this work and in particular in Figs.~\ref{fig:phases},~\ref{fig:mass-vs-y} and~\ref{fig:mass-vs-coupling}, we use the set of integrated
  Boltzmann equations from App.~\ref{sec:chem-dec}, unless specified otherwise. This
  method is computationally much faster than solving the unintegrated
  Boltzmann equations and reproduces the exact results to a good approximation.

\section{Bounds from cosmology and astrophysics}\label{sec:cosmo}
The parameter space of i2DM is constrained by several
observables in cosmology and astrophysics. In general, the presence of light new
particles with masses in the MeV-GeV range affects the thermal
history of the universe. Two main effects can be distinguished. First,
the presence of additional particles in the thermal bath changes the
evolution of the Hubble expansion and of the entropy density of the
universe. This may significantly affect the QCD phase transition, Big
Bang Nucleosynthesis (BBN), the Cosmic Microwave Background (CMB) or supernova cooling. Second, the annihilation or decay of new particles
injects energy into the thermal plasma by inducing excitations,
ionisation and heating. Such effects can modify BBN and the CMB compared to the predictions of the cosmological standard model.

In this section, we study the impact of i2DM dark fermions on cosmological and astrophysical observables. In Sec.~\ref{sec:qcd-pt}, we consider bounds on particles that freeze-out
around the QCD phase transition, at temperatures $T_{\rm QCD}\sim
200\,$MeV. In Sec.~\ref{sec:bbn},
we derive constraints from BBN around $T_{\rm BBN}\sim 0.1\,$MeV. In
Sec.~\ref{sec:cmb}, we discuss effects on the CMB around $T\sim
  1\,$eV, while in Sec.~\ref{sec:DNeff} we report on the constraints
  from the measurement of the effective number of neutrinos,
  $N_{\rm eff}$, at the CMB and BBN times. Finally, in
Sec.~\ref{sec:supernovae}, we consider constraints on dark sector
particles escaping supernovae.

\subsection{QCD phase transition}
\label{sec:qcd-pt}
If new particles freeze-out around the GeV scale, the QCD phase transition around $T_{\rm QCD}\sim 200\,$MeV affects the relic abundance~\cite{Steigman:2012nb}. The confinement of quarks and gluons into hadrons reduces the effective number of relativistic degrees of freedom contributing to the entropy density, $h_{\rm eff}$~\cite{Olive:1980dy}. Calculations of $h_{\rm eff}$ around the phase transition are subject to significant uncertainties, leading to variations of about ten percent in the relic abundance~\cite{Hindmarsh:2005ix,Laine:2006cp,Drees:2015exa}. However, the effect of the QCD phase transition on the thermal evolution of light new particles can be much larger than the mentioned uncertainties. In particular, it can affect the relic abundance of dark matter candidates that freeze out during or shortly after the phase transition.

In Sec.~\ref{sec:phases}, we have investigated the freeze-out dynamics of i2DM separately from effects of the QCD phase transition. If the freeze-out occurs in thermal equilibrium, dark matter candidates with masses
\begin{align}\label{eq:qcd-pt-bound}
    m_1\lesssim 1\,\text{GeV}
\end{align}
decouple from the bath around $T_{\rm WIMP} \approx m_1/x_{\rm WIMP} \ll T_{\rm
  QCD}$, late enough to neglect effects of the QCD phase transition on
dark matter decoupling~\cite{Drees:2015exa}. In i2DM this holds in the phases of coannihilation and partner annihilation. However, in the conversion phase, Eq.~(\ref{eq:qcd-pt-bound}) is not necessarily satisfied because chemical and kinetic equilibrium are not guaranteed until freeze-out, see
Sec.~\ref{sec:conversion}. In this case, the dark fermions decouple from chemical equilibrium at earlier times $x_{1,2}
\lesssim x_{\rm WIMP}$, which might be affected by the QCD phase transition if the corresponding decoupling temperature is similar to $ T_{\rm QCD}$. On the other hand, dark matter freeze-out is expected to be unaffected by the phase transition, because the dark states decouple at temperatures $T \lesssim T_{\rm WIMP}$. In any case, bounds from laboratory searches exclude i2DM dark matter candidates with masses near 1\,GeV, see Sec.~\ref{sec:colliders}. Even if non-equilibrium effects can change the decoupling times of the dark states, we do not expect effects from the QCD phase transition to affect the thermal history of dark matter candidates with masses well below the GeV scale.

\subsection{Big Bang Nucleosynthesis}
\label{sec:bbn}

The formation of primordial light nuclei starts around
$T_{\rm BBN} \sim 0.1\,$MeV. New physics can affect the nuclei
abundances in multiple ways. First of all, new dark particles can affect BBN by modifying the Hubble rate or the entropy
density of the universe. Second,
efficient annihilation of MeV-GeV-scale dark particles to electrons or photons may alter the rate at which light elements form~\cite{Depta:2019lbe}. Third, dark particles decaying to electrons or photons at later times can destroy the already formed primordial nuclei~\cite{Depta:2020zbh,Depta:2020mhj}. In i2DM, dark matter annihilation and dark partner decays around
$T \sim T_{\rm BBN}$ are not strong enough to cause observable effects. Late decays of dark partners with lifetimes $\tau_2 > t_{\rm
  BBN}$, however,
 can destroy the newly formed elements through photodisintegration and constrain parts of the i2DM parameter space. Below we will discuss photodisintegration in i2DM in detail and briefly argue why annihilation and decays during BBN are not efficient.

For annihilating dark matter during BBN, a lower mass bound of $m_{\rm DM} \gtrsim 10\,$MeV
 has been derived in~\cite{Depta:2019lbe}. This bound applies for vanilla WIMP annihilation with $\langle\sigma v\rangle \approx 10^{-26}-10^{-28}$
 cm$^3$/s. In the coannihilation and conversion phases of i2DM, however, the
 cross section for annihilations $\chi_1\chi_1\to \{e^+e^-,\gamma \gamma\}$ is much
 smaller than for vanilla WIMPs, resulting in
  weaker bounds on the dark matter mass. We expect these bounds to be superseded by constraints on $\Delta N_{\rm
   eff}$ from CMB measurements, see Sec.~\ref{sec:DNeff}.

On the other hand, decays $\chi_2 \to \chi_1 \ell^+\ell^-$ and $\chi_2 \to
\chi_1\gamma\gamma$ of dark partners with MeV-GeV masses and lifetimes $\tau_2 > 10^3\,$s can produce an electromagnetic cascade of photons with energies above the binding energy of light nuclei, consequently
disintegrating them.\footnote{For lifetimes $\tau_2\sim t_{\rm
    BBN}\sim 10^2$s, we have estimated from the analysis
  of~\cite{Depta:2020zbh} that no further constraints arise. This
  expectation is justified by comparing our i2DM predictions with BBN
  bounds on decaying dark scalars with a certain lifetime, shown
  in~Fig.~4 in~\cite{Depta:2020zbh}. Compared to~\cite{Depta:2020zbh},
  in i2DM we expect a smaller branching ratio to electrons and photons
 and softer spectra of the 3-body decays products; freeze-out
temperatures significantly lower than $10^{-2}\,$GeV for MeV-GeV particles; and an exponentially suppressed abundance of
the decaying dark partner $\chi_2$.} Disintegration is
efficient only if the emitted photons do not lose their energy too
rapidly before reaching the target nucleus. This condition is
satisfied if the energy of the photons lies below the di-electron
threshold, $E^{\rm th}_{e^+ e^-} \simeq
m_e^2/(22\,T)$~\cite{Hufnagel:2018bjp}.\footnote{Qualitatively this
  condition can be understood from the requirement that the
  center-of-mass energy of the injected photon and the thermal bath
  photon scaling as $E_\gamma E_{\gamma_{\rm th}}$ is of the order of
  $ m_e^2$, where $E_\gamma$ is the energy of the injected photon and
  $E_{\gamma_{\rm th}} \sim T$ is the average energy of a photon from
  thermal bath.} On the other hand, for photodisintegration to take place, $E^{\rm
  th}_{e^+ e^-}$ should lie well above the binding energy of light
elements, corresponding to temperatures below a few~keV. Photodisintegration is thus efficient for photon energies
\begin{align}
\mathcal{O}(\text{keV}) \lesssim E_\gamma \lesssim \frac{m_e^2}{22\,T}\,.
\end{align}
We calculate the effects of photodisintegration on the nuclei
abundances starting from a continuous electron spectrum originating
from $\chi_2 \to \chi_1 e^+e^-$ decays. We neglect loop-suppressed
direct photon production, $\chi_2 \to \chi_1 \gamma \gamma$, as well
as final-state radiation, resulting in a conservative bound on the electromagnetic flux obtained from partner decays~\cite{Forestell:2018txr}. To investigate photodisintegration for i2DM, we use the public code
ACROPOLIS~\cite{Depta:2020mhj}. The code calculates the modified
primordial abundances of light elements induced by
photodisintegration, accounting for the redistribution of the energy
injected by the decaying dark particle in the plasma.
   In particular, ACROPOLIS includes the
    exponential suppression in the photon spectrum for energies
    $E_{\gamma}>E^{th}_{e^+ e^-}$, which was neglected in previous
    studies,\footnote{Previous studies used the universal photon
      spectrum, which is an analytic approximation working very well
      for energies $E_{\gamma}<E^{th}_{e^+ e^-}$, but neglects photon
      with energies above this threshold.} up to energies
    $E_{\gamma}<E_0$, where $E_0$ is a model-dependent upper limit on the photon spectrum. We set $E_0$ to the maximum kinematically
    allowed energy. For the initial abundances of light nuclei, we use the
  \emph{Standard BBN} prediction extracted from the code AlterBBN~\cite{ARBEY20121822}. The input number density of
  $\chi_2$ is extracted from our system of Boltzmann equations.

  The resulting constraints
  from photodisintegration on partner decays in i2DM are shown in
  Fig.~\ref{fig:mass-vs-y}. Bounds on the displayed parameter space
  are only visible in the lower right corner of the left panel with
  $\tan \theta=10^{-4},\,\Delta=0.05$, far from the correct relic abundance
  (the green dotted line). At larger $\tan \theta$,
  photodisintegration is not efficient because the dark partners decay
  too early. The lifetime of $\chi_2$ also determines the upper edge
  of the excluded region. At dark matter masses, $m_1 \lesssim
  120\,$MeV, photodisintegration becomes inefficient due to the small
  absolute mass splitting between the dark fermions, $m_2-m_1$, which
  causes too soft decay products. For larger mass splittings
  $\Delta=0.1$, in the right panel, photodisintegration is sensitive
  to smaller dark matter masses. However, the lifetime of $\chi_2$ is
  generally smaller so that smaller couplings would be necessary for
  photodesintegration to take place. As a consequence, the excluded
  region lies below the plotted area in Fig.~\ref{fig:mass-vs-y} and
  BBN bounds are irrelevant for viable i2DM relics.  In summary, in i2DM
  photodisintegration excludes dark partners with lifetimes much
  larger than in cosmologically viable scenarios.

\subsection{Cosmic Microwave Background}
\label{sec:cmb}
Dark sector particles that annihilate or decay around the time of
recombination can affect the overall shape of the CMB black body
spectrum, as well as its temperature and the polarization anisotropy
spectra. Measurements of the CMB temperature and the polarization
anisotropy spectra set strong constraints on the annihilation cross
section of WIMP-like dark matter and also on decaying new particles
with lifetimes $\tau> 10^{13}\,$s. For particles with shorter
lifetimes, extra constraints can be obtained by studying deformations
of the blackbody spectrum before recombination, usually referred to as
spectral distortions. In addition, dark matter with couplings to
neutrinos, photons or electrons can shift the effective number of
neutrinos around the time of recombination $\Delta N_{\rm eff}(T_{\rm
  CMB})$, and affect the CMB anisotropies. We will discuss the
constraints arising from $\Delta N_{\rm eff}$
  in Sec.~\ref{sec:DNeff}.

Charged SM particles that can arise from decays or annihilations of dark particles can inject energy into the plasma in the form of heat. Heat injections at redshifts $z\lesssim 2\cdot 10^6$ induce spectral distortions in the CMB.
Measurements of spectral distortions are sensitive to dark particles with lifetimes $\tau\gtrsim 10^4\,$s.\footnote{In~\cite{Chluba:2013wsa} it has been shown that spectral
  distortions cannot set competitive bounds on dark matter
  annihilation compared to bounds from CMB anisotropies.} However, existing bounds
from the COBE-FIRAS experiment~\cite{Fixsen:1996nj} are largely superseded by the BBN bounds discussed in Sec.~\ref{sec:bbn}.\footnote{See also~\cite{Bolliet:2020ofj} for FIRAS/Planck constraints on decays
  into low-energy photons.}  Future CMB missions similar to PiXie could strengthen the
bounds on particles with lifetimes $\tau\gtrsim 10^4\,$s by up to two orders magnitude~\cite{Poulin:2015opa,Lucca:2019rxf}. In contrast, BBN bounds are
not expected to improve as much in the future.

Charged particles from decays or annihilations of dark sector
particles can also ionize the plasma. Ionization at redshifts
$z\lesssim 10^3$ modifies the CMB anisotropy spectra.\footnote{Any
  energy release into the plasma at redshifts earlier than $z\sim
  1400$ hardly affects the ionization history and has little impact on
  the CMB anisotropies~\cite{Chluba:2009uv,Bolliet:2020ofj}. As a
  result, MeV-GeV dark partners can only affect the CMB through
  spectral distortions.} This ionized fraction of the energy deposit
induces a broadening of the surface of last scattering; for instance,
it attenuates the CMB power spectrum on scales smaller than the width
of the surface~\cite{Padmanabhan:2005es}. Measurements by the Planck
collaboration constrain this effect and set strong upper limits on the
cross section for $s$-wave dark matter
annihilation~\cite{Planck:2018vyg}. In i2DM, dark matter annihilation
$\chi_1 \chi_1 \to f\bar{f}$ is suppressed well below these
limits.

Compared to ionization effects on the CMB anisotropies, searches for dark matter annihilation in indirect detection experiments impose much weaker bounds in the MeV-GeV mass range~\cite{Leane:2018kjk,e-ASTROGAM:2017pxr}. Future missions such as e-astrogram~\cite{e-ASTROGAM:2017pxr} can provide competitive bounds on the annihilation cross section, which however are still far above the suppressed annihilation rates in i2DM.


\subsection{Effective number of neutrinos $ N_{\rm eff}$}
\label{sec:DNeff}
As mentioned above, stable dark matter coupled to neutrinos,
photons or electrons can change the effective number of neutrinos by $\Delta N_{\rm eff}$.
 This modification can affect the
abundance of light nuclei (set at $T_{\rm BBN } \sim 0.1$ MeV) and the
CMB anisotropies (set around $T_{\rm CMB } \sim 0.4$ eV). Both
measurements can thus set bounds on $\Delta N_{\rm eff}$.

Efficient dark matter scattering with the electromagnetic or neutrino bath can induce an entropy transfer to these species after neutrino decoupling
(at $T_\nu \sim$ MeV), thus modifying the neutrino-to-photon temperature ratio
compared to the SM prediction. In practice, this information is
encapsulated in
 $\Delta N_{\rm eff}$. For dark matter that only couples to
neutrinos, the entropy transfer reheats the neutrino bath and
induces a positive shift $\Delta N_{\rm eff} > 0$. For dark matter coupling to either
electrons or photons, the electromagnetic bath gets reheated, which induces a negative shift $\Delta N_{\rm eff} < 0$.

To estimate the effect of light dark particles on $N_{\rm eff}$,
 we follow the detailed analysis of~\cite{Escudero:2018mvt}, which uses the constraint on $\Delta N_{\rm eff}$ by the Planck collaboration~\cite{Planck:2018vyg},
  \begin{align}
 N_{\rm eff}(T_{\rm CMB})=2.99^{+0.34}_{-0.33}\qquad \text{at } 95\%\,\text{CL}.
 \end{align}
From this constraint, the authors of~\cite{Escudero:2018mvt} derive a  lower bound on the mass of a Dirac fermion dark matter candidate that couples efficiently to electrons, finding
\begin{equation}
m_{\rm DM} > 9.2\, {\rm MeV}\qquad \text{at } 95\%\,\text{CL}.
  \label{eq:DNeffbound}
\end{equation}
This result agrees with the estimates in~\cite{Boehm:2013jpa,Depta:2019lbe}, for
instance. We emphasize that a modification of $N_{\rm eff}$ can only
be observed in CMB data if~\cite{Boehm:2013jpa}
\begin{itemize}
\item[(i)] the dark matter is in
kinetic equilibrium with either electrons, photons or neutrinos
 at temperatures above and below $T_\nu \sim$~MeV;
\item[(ii)] the dark matter becomes
non-relativistic at temperatures $T \lesssim T_\nu$, typically for
masses below a few tens of MeV.
\end{itemize}
 As a result, in i2DM the bound from Eq.~(\ref{eq:DNeffbound}) only
 holds in regions (A) and (B), defined in Eqs.~(\ref{eq:equ}) and
 (\ref{eq:chemdec}), where the dark fermions are in kinetic
 equilibrium prior to freeze-out.  For simplicity, we restrict
 ourselves to dark matter candidates with $m_1 > 10\,$MeV in all three
 regions.

\subsection{Supernova cooling}
\label{sec:supernovae}
Feebly interacting dark-sector particles can be produced in proto-neutron stars and freely escape them. The energy carried by the dark particles speeds up the supernova cooling and therefore can constrain models with light dark sectors. In i2DM, dark fermions are mostly produced through decays of dark photons produced in bremsstrahlung during neutron-proton collisions or directly in collisions of light SM fermions~\cite{Chang:2018rso}. Existing bounds on the cooling of the supernova SN1987A~\cite{Kamiokande-II:1987idp,PhysRevLett.58.1494}
 constrain the i2DM parameter space only at very small couplings that are irrelevant for our purposes. Based on the results for iDM in~\cite{Chang:2018rso}, we estimate that in i2DM supernova cooling constrains dark couplings up to at most $y \approx 10^{-13}$ for masses $m_1\lesssim 200\,$MeV; a parameter space where the dark matter relics would be overabundant. The bounds could potentially be even weaker if different core-collapse simulations were used~\cite{Bar:2019ifz}. Valid i2DM relics are thus not subject to bounds from supernova cooling.

\section{Laboratory searches}\label{sec:colliders}
In this section we discuss the phenomenology of inelastic dark matter in direct detection experiments, at particle colliders and at fixed-target experiments. Despite the original motivation of inelastic dark matter to evade direct detection, recent investigations reveal sensitivity to certain scenarios of iDM. Colliders and fixed-target experiments set strong bounds on the parameter space of iDM. We show that i2DM can evade some of these bounds, but can be conclusively tested at future experiments. Indirect detection searches are not sensitive to i2DM, because the pair-annihilation rate of dark matter is suppressed by small kinetic mixing and small dark fermion mixing, well below the reach of current and projected future experiments.

\subsection{Direct detection}
\label{sec:direct-detection}
In scenarios of feebly interacting inelastic dark matter, elastic
scattering $\chi_1 X\to \chi_1 X$ off nucleons or electrons,
  $X=n$ or $e$, is suppressed below the sensitivity of current
direct detection experiments. In the MeV-GeV mass range, the recoil energy in nucleon scattering is typically too small to be observed, but electron recoils are a promising road to detection~\cite{Ema:2018bih}. The strongest current bound on dark matter-electron scattering by Xenon 1T lies around $\sigma(\chi_1 e\to \chi_1 e) \sim 10^{-40}$ cm$^2$~\cite{XENON:2019gfn} for $m_1\approx 100\,$MeV. For comparison, in i2DM with $m_1= 100\,$MeV, $m_{A'}= 3  m_1$ and large mixings $\sin \theta = 0.1$ and $\epsilon = 0.02$, the predicted cross section is $\sigma(\chi_1 e\to \chi_1 e)=6\times 10^{-41}\,{\rm cm}^2$. Future direct detection experiments might reach a higher sensitivity to dark matter-electron scattering. Whether they can reach i2DM sensitivity will depend on the progress in collider searches (see Sec.~\ref{sec:idm-colliders}), which can probe kinetic mixing down to $\epsilon \approx 10^{-3}$ and potentially suppress the target cross section as $\sigma\sim \epsilon^2$.

If the mass splitting $\Delta$ between the
dark states is smaller than a few 100 keV,
 inelastic up-scattering $\chi_1 n \to \chi_2 n$ can produce an observable nuclear recoil signal, provided that the threshold of the experiment is low enough to detect the recoil energy~\cite{Tucker-Smith:2001myb,Bramante:2016rdh}.

For larger mass splitting, up-scattering can be observable if dark matter is accelerated through interactions with cosmic rays in the atmosphere around the earth~\cite{Bringmann:2018cvk}: The dark matter candidate $\chi_1$ scatters inelastically off cosmic rays, mostly protons, via $\chi_1 p \to \chi_2 p$. Subsequent decays $\chi_2 \to \chi_1 f\bar{f}$ produce a relativistic component of dark matter, which passes the energy threshold for $\chi_1 n \to \chi_2 n$ scattering in experiments~\cite{Bell:2021xff}.\footnote{Up-scattering can also occur in the Sun or the Earth~\cite{Baryakhtar:2020rwy,Emken:2021vmf}, but is not efficient enough to be probed in current direct detection experiments for the dark matter scenarios considered in this work.}

Dark partners produced from cosmic-ray reactions can induce nuclear recoils via down-scattering $\chi_2 n\to \chi_1 n$, provided that their lifetime is long enough to reach the experiment~\cite{Graham:2010ca,Bell:2021xff,CarrilloGonzalez:2021lxm}. Down-scattering off electrons $\chi_2 e\to \chi_1 e$ is an interesting alternative, which can also address the current excess of electron recoils at Xenon1T~\cite{Aboubrahim:2020iwb}. For inelastic dark matter, efficient down-scattering only occurs if the lifetime is longer than several years~\cite{Bell:2021xff}, corresponding to mass splittings much smaller than those considered in this work.

For i2DM, we expect that the rates for up-scattering and down-scattering are generally suppressed compared to iDM, due to the smaller $A'\chi_1\chi_2$ coupling proportional to $\tan\theta$.
 On the other hand, in i2DM scenarios with small mass splitting, dark partners produced from cosmic-ray up-scattering are long-lived enough so that a substantial fraction of them could reach the detector before decaying. In this case, elastic scattering via $\chi_2 n \to \chi_2 n$ should dominate and leave an interesting characteristic signature of i2DM. In such scenarios, also down-scattering is expected. If the dark states are heavy and compressed enough to induce an observable nuclear recoil, up-scattering of i2DM off cosmic rays followed by elastic scattering off nuclei in the detector could be directly probed at experiments with a low energy threshold, cf. Ref.~\cite{CarrilloGonzalez:2021lxm}. A dedicated analysis of i2DM at direct detection experiments goes beyond the scope of this work, but is a promising direction for future research.

\subsection{Electroweak precision observables}
\label{sec:ewpo}
A general bound on kinetic mixing of a dark photon is obtained from electroweak precision tests. In electroweak observables measured at LEP, Tevatron and the LHC, kinetic mixing modifies the $Z$ boson's mass and couplings to SM fermions at $\mathcal{O}(\epsilon^2)$. For dark photons with masses well below the $Z$ resonance, a global fit to electroweak precision data yields a 95\% CL upper bound of~\cite{Curtin:2014cca}
\begin{align}\label{eq:ewp-bound}
\epsilon \lesssim 0.02 \qquad \text{for}\qquad  m_{A'} \lesssim 10\,\text{GeV};
\end{align}
stronger bounds apply for dark photons with masses closer to the $Z$ pole.
For $m_{A'} \lesssim 10\,$GeV, slightly stronger bounds have also been obtained from $e^\pm$ scattering off protons at HERA~\cite{Kribs:2020vyk}.

\subsection{Collider searches}
\label{sec:idm-colliders}
At $e^+e^-$ colliders, dark fermions coupling via a dark photon can be produced via three main processes:
\begin{align}
    e^+e^- & \to A_D \gamma \to \chi_1 \chi_1 \gamma\\\nonumber
    e^+e^- & \to A_D \gamma \to \chi_1 \chi_2 \gamma \to \chi_1 \chi_1 \ell^+\ell^- \gamma\qquad \qquad \text{ (iDM)}\\\nonumber
    e^+e^- & \to A_D \gamma \to \chi_2 \chi_2 \gamma \to \chi_1 \ell^+\ell^- \chi_1 \ell^+\ell^- \gamma \qquad \text{(i2DM)}.
\end{align}
In models of inelastic dark matter, the first process is suppressed by construction. The second process dominates in iDM scenarios, which typically rely on the coupling of the dark photon to $\chi_1$ and $\chi_2$. The third process is characteristic of i2DM, since the coupling $A_D\chi_1\chi_2$ is suppressed by $\tan\theta$ and the dark photon mostly decays via $A_D \to \chi_2\chi_2$, see Eq.~\eqref{eq:brs}. Dark photon decays into SM fermions are suppressed as $\epsilon^2 \alpha/\alpha_D$, so that resonance searches at BaBar~\cite{BaBar:2014zli} and LHCb~\cite{LHCb:2017trq,LHCb:2020ysn} are not sensitive to the scenarios investigated in this work.

Collider signals of inelastic dark matter depend on whether the dark partners decay within or outside the detector. Below the hadronic threshold, the lifetime of $\chi_2$ is mostly determined by $\chi_2\to \chi_1 \ell^+\ell^-$ decays, which strongly depend on  $\epsilon\tan\theta$ and the mass splitting $\Delta$, see Eq.~\eqref{eq:decay-rate}. If one or two dark partners decay within the detector, the signature consists of one or two prompt or displaced vertices of charged leptons, in association with a photon and missing energy. In iDM, the phenomenology of this signature has been investigated in detail for the Belle II experiment~\cite{Duerr:2019dmv,Mohlabeng:2019vrz,Kang:2021oes}. For sufficiently large mass splitting $\Delta$, Belle II will be able to probe scenarios of iDM in the GeV range.

For a smaller mass splitting, the decay products of $\chi_2$ are too soft to be detected, leading to a signal with a photon and missing energy. The same signature is expected if $\chi_2$ decays outside the detector. A search for mono-photon signals at BaBar has set an upper bound on the kinetic mixing of \emph{invisible} dark photons~\cite{BaBar:2017tiz},
\begin{align}\label{eq:mono-photon-bound}
\epsilon \lesssim 10^{-3}\qquad \text{for}\qquad  m_{A'} \lesssim 5\,\text{GeV}.
\end{align}
In iDM, this bound excludes most of the parameter space for dark matter candidates below the GeV scale. A similar search at Belle II can probe even smaller dark sector couplings and thereby improve the sensitivity to iDM~\cite{Duerr:2019dmv}.

\begin{figure}[t!]
	\centering
	\includegraphics[scale=0.56]{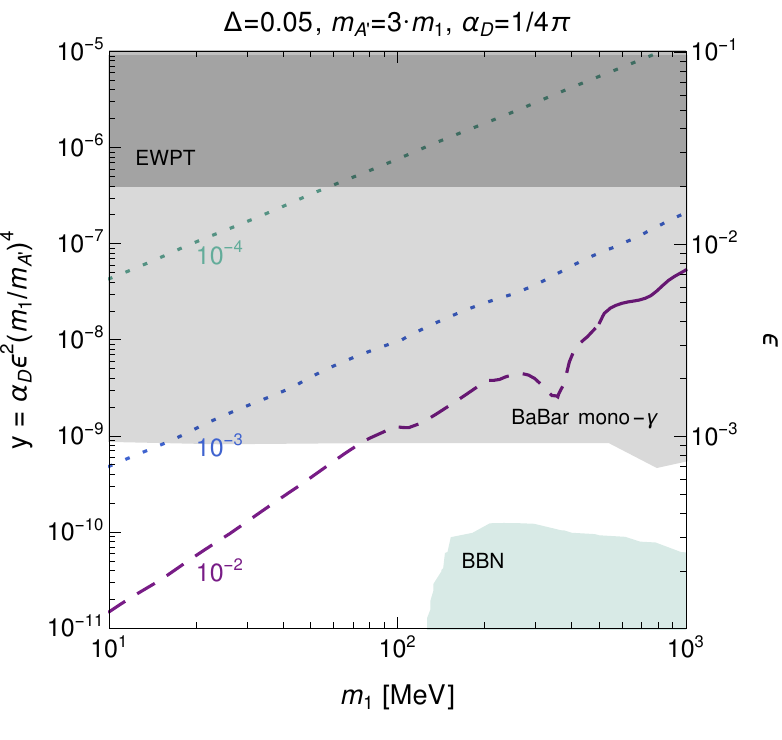}
	\includegraphics[scale=0.56]{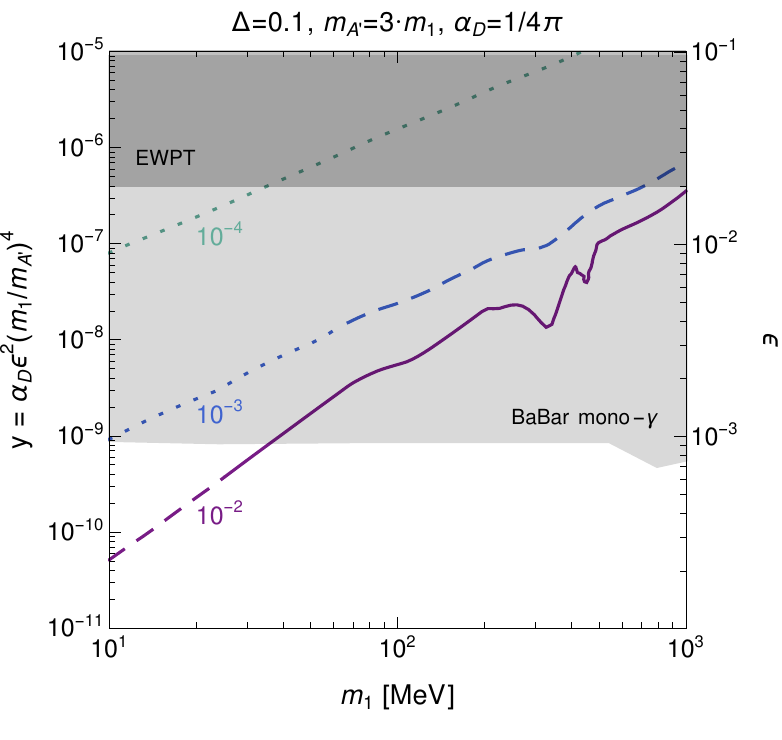}
	\caption{Collider bounds in the $(m_1,y)$ plane of the
          i2DM. The dark gray area is excluded by electroweak
          precision observables bound of Eq.~(\ref{eq:ewp-bound})
          while the light gray area is excluded by BaBar's mono-photon
          search, see Sec.~\ref{sec:idm-colliders}. The green
            area (BBN) is excluded for $\theta=10^{-4}$ due to
            late photodisintegration of light nuclei, see Sec.~\ref{sec:bbn}. Colored curves
          correspond to contours of $\Omega_\chi h^2=0.12$ for fixed
          values of $\tan\theta=10^{-2}$ (purple), $10^{-3}$ (blue)
          and $10^{-4}$ (green) for fixed relative mass splitting
          $\Delta=0.05$ (right) and 0.1 (left). On the curves, the
            three regions introduced in Sec.~\ref{sec:phases} are
            indicated as plain (A), dashed (B), and dotted (C).
\label{fig:mass-vs-y}}
\end{figure}

In i2DM, the dark photon decays close to its production point due to efficient $A' \to \chi_2\chi_2$ decays. However, the dark partners have larger decay lengths than in iDM, because $\chi_2\to \chi_1 f\bar{f}$ decays are suppressed by $\tan^2\theta$, see Eq.~\eqref{eq:decay-rate}. Therefore the dark photon does not leave any trace in the detector and the bound of Eq.~\eqref{eq:mono-photon-bound} from BaBar's mono-photon search applies. In Fig.~\ref{fig:mass-vs-y}, we display the bounds on kinetic mixing from collider searches in the parameter space of i2DM for $\Delta=0.05$ (left) and 0.1 (right). The observed relic abundance is obtained along the contours for fixed values of the dark fermion mixing $\tan\theta=10^{-2}$, $10^{-3}$ and $10^{-4}$. Small values of $\tan\theta$ need to be compensated by a larger effective interaction strength $y$ to avoid overabundance in the coscattering regime, see Fig.~\ref{fig:phases}. BaBar's mono-photon search translates into a strong upper bound on $y$. In the remaining parameter space, the dark matter abundance is set by partner annihilation in region (A) (plain) and, for small masses, by coscattering in regions (B) (dashed) or (C) (dotted). Departures from kinetic equilibrium, occurring in region (C), only occur in a small region of parameter space. Most viable i2DM candidates are therefore in kinetic equilibrium before freeze-out.
%

\subsection{Bounds from fixed-target experiments}
\label{sec:fixed-target}
Fixed-target experiments with a large separation of the particle source and the detector are particularly sensitive to particles with a long decay length. Searches for long-lived particles with sub-GeV masses have
been performed at various fixed-target experiments and have been
reinterpreted for iDM scenarios, for instance in
Refs.~\cite{Gninenko:2012eq,Izaguirre:2017bqb,Tsai:2019buq}. In general, a beam of
  particles is dumped on a target material, producing light
  mesons such as pions or kaons. In iDM and i2DM, dark photons can be produced either in meson decays, through the Primakoff process or from bremsstrahlung. Subsequently the dark photons decay into pairs of dark fermions $\chi_1$ and/or $\chi_2$.

  Depending on the model parameters, inelastic dark matter can be detected via three signatures: (displaced) decays of dark partners, $\chi_2 \to \chi_1 \ell^+\ell^-$; (up-)scattering of dark fermions off the detector material, $\chi_i N \to \chi_j N$; or missing energy from dark fermions that are stable at the scales of the experiment. The relative sensitivity to each signature depends on the lifetime of the dark partner, as well as on the experimental setup: short-lived dark partners are mostly observed through decays, while long-lived partners scatter inside the detector material or decay after passing through the detector. In what follows, we discuss the relevant signatures for i2DM and derive bounds from searches at fixed-target experiments.

\paragraph{Partner decays} For partner decays, the expected event rate in the far detector is given by\footnote{Here we neglect dark partners produced via upscattering $\chi_1 \to \chi_2$.}
\begin{align}
N_{\rm dec} \approx N_{A'}\Big(\mathcal{B}(A'\to \chi_1\chi_2) + 2 \mathcal{B}(A'\to \chi_2\chi_2)\Big) \frac{1}{N_2}\sum_{k=1}^{N_2}\,P_{\rm dec}(d_k)\,,
\end{align}
where $N_{A'} \propto \epsilon^2$ is the total number of  dark photons produced in a given experiment, and $N_2$ is the total number of $\chi_2$ states resulting from $N_{A'}$ dark photon decays. The branching ratios are $\mathcal{B}(A'\to \chi_1\chi_2)\approx 1$ for iDM and $\mathcal{B}(A'\to \chi_2\chi_2)\approx 1$ for i2DM with $\tan\theta \ll 1$.\footnote{In our numerical analysis of i2DM, we neglect $A'\to \chi_1\chi_2$ decays.} Finally, $P_{\rm dec}(d_k)$ is the probability to detect the decay products of particle $k$ with decay length $d_k$. The decay length $d_k = (\beta\gamma)_k c\tau_2$ depends on the boost, $(\beta\gamma)_k$, and on the lifetime, $\tau_2$, of the dark partner. For decay lengths longer than the baseline, the decay probability scales as $P_{\rm dec}(d) \propto 1/d$. In this regime, the expected event rates for iDM and i2DM depend on the model parameters as
\begin{align}\label{eq:partner-decays}
N_{\rm dec}^{\rm iDM} \propto \epsilon^2 y \propto \epsilon^4 \alpha_D\,,\qquad N_{\rm dec}^{\rm i2DM} \approx 2 \tan^2\theta \cos^4\theta\, N_{\rm dec}^{\rm iDM}\,.
\end{align}
In i2DM, the lifetime of the dark partner scales as $\tau_2 \sim 1/\tan^{2}\theta$. For small $\tan\theta$, the dark partner tends to decay after passing the detector, which reduces the event rate. The factor of 2 accounts for the two dark partners produced in $A'\to \chi_2\chi_2$ decays. For fixed parameters $\alpha_D,\,m_1,\,m_{A'},\,\Delta$, a bound on $\epsilon$ obtained from searches for dark partner decays in iDM translates into a bound on $\epsilon \cdot (2\tan^2\theta\cos^4\theta)^{1/4}$ in i2DM.

\begin{figure}[ht!]
  \centering
  \includegraphics[scale=0.57]{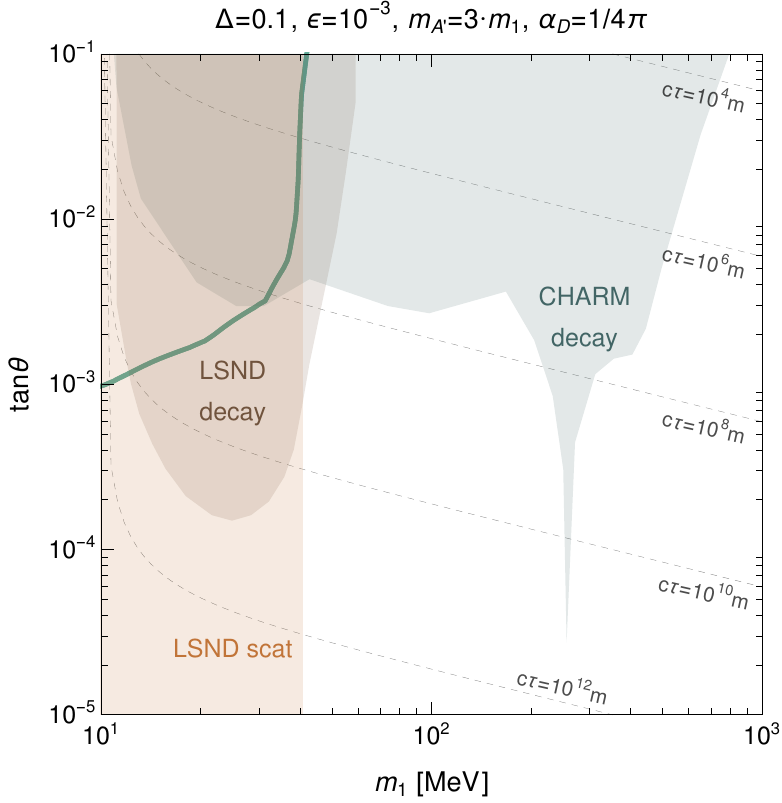}\hspace*{0.2cm}
  \includegraphics[scale=0.57]{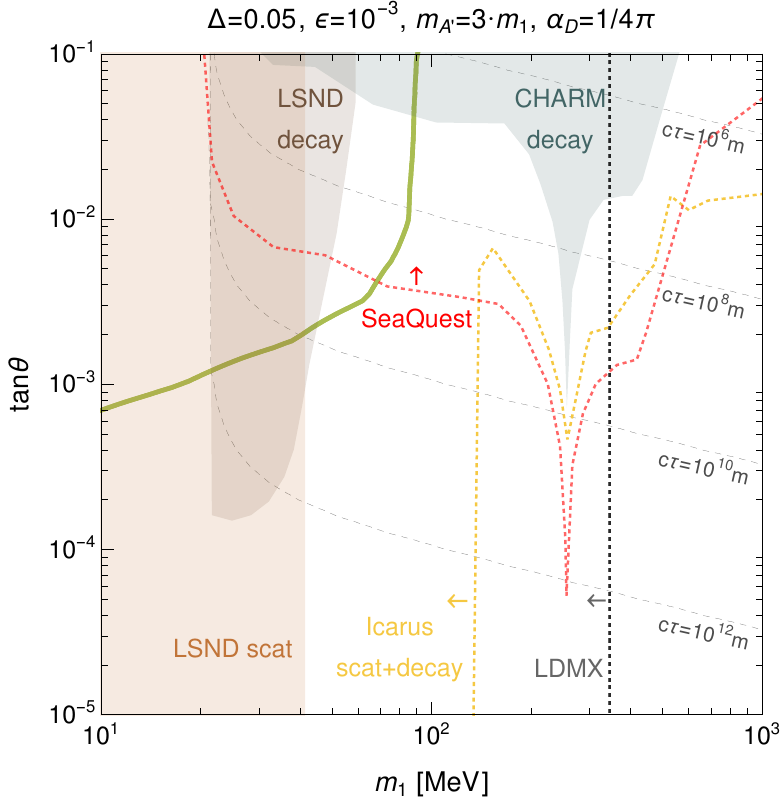}\\
  \vspace*{0.7cm}
  \includegraphics[scale=0.57]{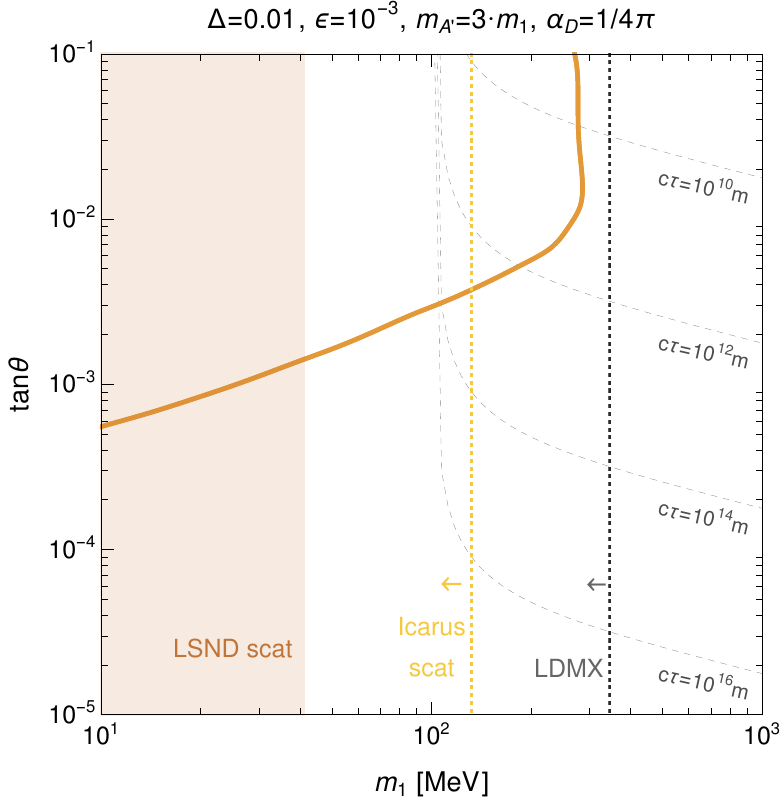}
  \caption{Bounds on i2DM from fixed-target experiments as a function of the dark fermion mixing $\theta$ and the dark matter mass $m_1$. Shown are three scenarios with a relative mass splitting $\Delta = 0.1$, $0.05$, $0.01$, and fixed dark-photon mixing $\epsilon= 10^{-3}$, dark-photon mass $m_{A'}= 3 m_1$ and dark coupling strength $\alpha_D=1/(4\pi)$. The observed relic abundance is obtained along the colored contours. The dashed contours indicate the proper
      decay length of the dark partner. For $m_1 < 2m_e/\Delta$, the dark partner can only decay into photons or neutrinos, resulting in a very long lifetime. The bounds have been obtained by rescaling
      dedicated analyses for iDM.  We show existing bounds from CHARM (grey areas) and LSND (red areas). The projected sensitivity of future experiments ICARUS (yellow), SeaQuest (red) and LDMX (grey) is illustrated by dotted lines; the arrow indicates the direction in parameter space that will be probed. All scenarios evade existing bounds from a mono-photon search at BaBar and electroweak precision tests (EWPT). See text for details.
\label{fig:mass-vs-coupling}}
\end{figure}

\paragraph{Decays at CHARM} Strong bounds on long-lived dark particles decaying into electrons have been set at the neutrino experiment CHARM. At CHARM, dark fermions can be efficiently produced from $\pi^0$ or $\eta$ meson decays, $\pi^0(\eta) \to \gamma A' \to \gamma \chi_i\chi_j$. Null results of a beam-dump search for heavy neutrinos decaying into electron pairs~\cite{CHARM:1983ayi} have been reinterpreted for $\chi_2\to \chi_1 e^+ e^-$ decays in iDM~\cite{Tsai:2019buq}. In Fig.~\ref{fig:mass-vs-coupling}, we show the resulting bounds in the parameter space of i2DM, using Eq.~\eqref{eq:partner-decays} to rescale the predicted event rates. The three scenarios are distinguished by the mass splitting $\Delta$, while we have fixed $\epsilon = 10^{-3}$ to evade the collider bounds from Sec.~\ref{sec:idm-colliders}.
The observed relic abundance is obtained along the colored contours. The lifetime of $\chi_2$ varies strongly with the mass splitting, $\tau_2 \propto 1/\Delta^{5}$, see Eq.~\eqref{eq:decay-rate}. For $\Delta = 0.1$, most of the dark partners decay within the considered decay volume and the search is sensitive to even small mixing $\tan\theta$. For $\Delta = 0.05$, the sensitivity decreases due to the longer lifetime and softer $e^+e^-$ momenta, which are less likely to pass the analysis cuts. For $\Delta = 0.01$, the dark partner is essentially stable compared to the length of the decay volume and the search becomes insensitive to i2DM.

\paragraph{Decays at LSND} For light dark sectors, even stronger bounds have been obtained from the neutrino experiment LSND. At LSND, dark fermions can be abundantly produced from pion decays via $\pi^0 \to \gamma A' \to \gamma \chi_i\chi_j$. In Ref.~\cite{Izaguirre:2017bqb}, LSND data has been interpreted in terms of $\chi_2 \to \chi_1 e^+ e^-$ decays in iDM, under the conservative assumption that the $e^+e^-$ pair is not resolved in the calorimeter and can mimic elastic neutrino-electron scattering. In Fig.~\ref{fig:mass-vs-coupling}, we show the resulting bounds rescaled for i2DM (labelled `LSND decay'). LSND is very sensitive to dark partners with $m_2 < m_{\pi^0}/2$. The lower cutoff is determined by the kinematic threshold for $\chi_2\to \chi_1 e^+e^-$ decays. The sensitivity could be improved with a dedicated analysis of three-body decays, rather than a re-interpretation of neutrino-electron scattering.

\paragraph{Dark fermion scattering} In addition to decays, long-lived dark fermions can be detected through up-scattering $\chi_1 \to \chi_2$, down-scattering $\chi_2\to \chi_1$, or elastic scattering $\chi_2 \to\chi_2$ inside the detector. We neglect up-scattering, which is typically sub-dominant for suppressed $\chi_1-\chi_2$ couplings.
The expected scattering rate is then given by
\begin{align}
N_{\rm scat} \approx N_{A'}\Big(\mathcal{B}(A'\to \chi_1\chi_2) + 2 \mathcal{B}(A'\to \chi_2\chi_2)\Big) \frac{1}{N_2}\sum_{k=1}^{N_2}\,P_{\rm scat}(d_k)\,,
\end{align}
where $P_{\rm scat}(d_k)$ is the probability for particle $k$ to scatter off the material inside the detector, with $d_k \to \infty$ for $\chi = \chi_1$, and $N_2$ is the number of produced dark partners. Neutrino experiments are particularly sensitive to dark fermion scattering, which mimics neutrino-electron scattering. Among various experiments, LSND sets the currently strongest bounds on inelastic dark matter. For iDM and sub-GeV masses, the relevant process is down-scattering $\chi_2 N \to \chi_1 N$~\cite{Izaguirre:2017bqb}. For i2DM, elastic scattering $\chi_2 \to \chi_2$ dominates for $\tan\theta \ll 1$. The respective event rates scale as
\begin{align}
N_{\rm scat}^{\rm iDM} \propto \epsilon^4\alpha_D\,,\qquad N_{\rm scat}^{\rm i2DM} \approx 2 \cos^4\theta\, N_{\rm scat}^{\rm iDM}
\end{align}
for decay lengths $d_k$ larger than the distance between the target and the detector.

\paragraph{Scattering at LSND} The LSND bounds on scattering are derived from the same analysis as for partner decays. Again, we translate the results for iDM from Ref.~\cite{Izaguirre:2017bqb} to i2DM, shown in Fig.~\ref{fig:mass-vs-coupling} as `LSND scat'. For $\tan\theta \lesssim 0.3$, the bounds are insensitive to dark fermion mixing and exclude small dark matter masses.
  The bounds disappear for small kinetic mixing $\epsilon \lesssim 10^{-4}$, where LSND loses its sensitivity due to the low dark photon production rate.

It is interesting to compare these results for i2DM with iDM. For $\tan\theta = 1$, the $A'\chi_1\chi_2$ interaction strength is similar in iDM and i2DM. For the benchmark scenarios shown in Fig.~\ref{fig:mass-vs-coupling}, iDM is excluded by CHARM and LSND, unless partner decays into electrons are kinematically forbidden. Viable scenarios of sub-GeV iDM require a stronger interaction strength $y$ for efficient coannihilation, while keeping the kinetic mixing $\epsilon$ small to evade bounds from mono-photon searches~\cite{Batell:2021ooj}, see Sec.~\ref{sec:idm-colliders}.
 In turn, i2DM scenarios with small $\tan\theta$ can evade the CHARM bounds due to the longer lifetime of the dark partners. In this regime, the relic abundance is set by partner annihilation and/or coscattering. In summary, null searches at current fixed-target experiments are a severe challenge for iDM, while i2DM is a viable option due to the impact of partners on dark matter freeze-out.

\subsection{Prospects of future fixed-target experiments}
\label{sec:fixed-target-future}
As we discussed in Sec.~\ref{sec:fixed-target}, existing fixed-target experiments are very sensitive to sub-GeV inelastic dark matter. However, in the multi-GeV range i2DM scenarios with a compressed dark sector currently escape detection. In order to fully explore the parameter space of i2DM, we study the discovery potential of proposed searches at fixed-target experiments, which could be realized in the near future. While a number of experiments can be sensitive to i2DM, here we focus a few promising proposals.

\paragraph{SBN}
The Short-Baseline Neutrino (SBN) program at Fermilab~\cite{Machado:2019oxb} is a planned facility to probe neutrinos and light dark sectors. The facility uses the Booster 8 GeV proton beam hitting a beryllium target. Three detectors are placed downstream of the target at varying distances. The SBND detector is located at around 110\,m, while the experiments MicroBooNE and ICARUS are placed further away, at 470\,m and 600\,m respectively.

At SBN, similarly to other fixed-target experiments discussed in Sec.~\ref{sec:fixed-target}, dark partners can be produced from decaying dark photons, which are created in the interaction of the proton beam with the target. The SBN detectors can be used to search for $\chi_2 \to \chi_1 \ell^+\ell^-$ decays or scattering $\chi_i \rightarrow \chi_j$ off the detector material.

Dark sector searches at neutrino experiments inevitably feature a large neutrino background. At SBN, there are two proposals to reduce this background. One option is to deflect the proton beam around the target into an iron absorber placed 50\,m downstream, which has previously been done at MiniBooNE to study light dark sectors \cite{MiniBooNE:2017nqe,MiniBooNEDM:2018cxm}. This mode is referred to as ``off-target''. The second option is to use the NuMI 120\,GeV proton beam, impacting on a graphite target. ICARUS and MicroBooNE are placed at angles of 6 and 8 degrees against the NuMI beam direction. This option is known as ``off-axis''. Both off-target and off-axis options have been studied for iDM~\cite{Batell:2021ooj}, where SBND has the best sensitivity in the off-target mode, while ICARUS can set the strongest bound in the off-axis mode. In Fig.~\ref{fig:mass-vs-coupling}, we show our reinterpretation of these predictions for i2DM, following the procedure described in Sec.~\ref{sec:fixed-target}. We present the results for ICARUS, assuming that the off-axis mode can be realized with much less technological effort than the off-target option. Compared to existing fixed-target experiments, ICARUS is substantially more sensitive to i2DM, provided that the dark partners are sufficiently long-lived to induce enough signal in the detector.

\paragraph{SeaQuest}
Originally developed to study the sea quark content of the proton with a 120 GeV proton beam and various targets, the Fermilab experiment SeaQuest has a good potential to probe dark sectors \cite{Gardner:2015wea,Berlin:2018pwi}. It is already equipped with a displaced muon trigger to study exotic long-lived particles decaying to muons, and could be supplemented by an electromagnetic calorimeter to also probe electron signals. In Ref.~\cite{Berlin:2018pwi}, the sensitivity of SeaQuest to dark partner decays in iDM has been studied for three different decay volumes. In Fig.~\ref{fig:mass-vs-coupling}, we show the corresponding predictions for i2DM for the largest possible decay volume. Compared with CHARM and ICARUS, SeaQuest can improve the sensitivity to i2DM for dark matter with masses near the $\eta$ resonance. As discussed in Ref.~\cite{Berlin:2018pwi}, the reach of SeaQuest could be further enhanced by running the experiment without the magnet, which however would require a dedicated analysis of the experimental setup.

\paragraph{LDMX}
The proposed Light Dark Matter eXperiment (LDMX)~\cite{LDMX:2018cma} is an electron beam-dump experiment designed primarily for probing light dark matter models. Its search strategy relies on dark sector particles being produced in the beam dump that escape the detector, which extends up to about $1\,$m downstream from the target. This gives rise to a signature of missing energy. All charged particles in an event are vetoed, except for the soft remnant of the incoming electron. A signal of inelastic dark matter is detected if a substantial fraction of dark partners decay \emph{after} passing through the detector. The LDMX collaboration has investigated the projected sensitivity to many MeV-GeV dark sector models, including iDM~\cite{Berlin:2018bsc}. In Fig.~\ref{fig:mass-vs-coupling}, we show our interpretation for i2DM for the most conservative design option. LDMX is well suited to probe i2DM scenarios with very small mass splitting $\Delta$, which are difficult to detect in experiments that detect the decay products of the dark partner.\\

We summarize our projections for future fixed-target experiments for the three i2DM benchmarks in Fig.~\ref{fig:mass-vs-coupling}. For $\Delta=0.1$, the dark matter target is already probed by existing experiments. We therefore do not show projections for future experiments, but note that they could be sensitive to other regions of the parameter space. For $\Delta=0.05$, SeaQuest alone can improve the sensitivity to dark partner decays, but cannot fully probe the dark matter target. ICARUS can complement SeaQuest by also detecting dark fermion scattering, which is predominant for small dark matter masses. LDMX, searching for missing energy, can extend the reach of ICARUS at larger masses and long lifetimes. Either ICARUS or LDMX could conclusively probe this scenario. For even smaller mass splitting $\Delta=0.01$, dark partner decays cannot be observed due to the long lifetime and the softness of the SM decay products. The sensitivity of ICARUS and LDMX through scattering and missing energy, however, is kept and allows to conclusively probe the dark matter scenario. Larger dark photon masses $m_{A'} > 3 m_1$ or smaller kinetic mixing $\epsilon$ reduces the rate of produced dark photons and thus the sensitivity of any  experiment. For smaller dark couplings $\alpha_D$, the lifetime of the dark partners is enhanced and dark partner decays close behind the target are less abundant. In this case, scattering and missing energy are more promising signals, especially for small dark matter masses. All in all, fixed-target experiments have a high potential to conclusively test i2DM in the near future, provided that they are built and successfully run.

\section{Conclusions and outlook}\label{sec:conclusions}
\noindent
In this work we have introduced a new model for feebly coupling dark matter, called inelastic Dirac Dark Matter. Compared to the widely studied model of inelastic Dark Matter with Majorana fermions, i2DM has a different thermal history and is less constrained by current laboratory searches. The main difference is due to the variable interaction strength of the dark matter candidate $\chi_1$, parametrized by a mass mixing $\theta$ with its dark partner $\chi_2$.

At small mass mixing, the dark matter candidate decouples from the SM bath before freeze-out and the relic dark matter abundance cannot be set by dark matter annihilation, coannihilation  or partner annihilation anymore. Instead, coscattering and decay processes are crucial to explain the observed abundance even for tiny dark matter interactions $y$ with the thermal bath. At such feeble couplings, dark matter can decouple from chemical and even kinetic equilibrium before freeze-out. We have computed the resulting effects on the relic abundance by numerically solving a coupled set of Boltzmann equations for the time evolution of the dark fermions.  For particles in the multi-MeV range, we find that the relic abundance can be obtained with interactions as feeble as $y \gtrsim 10^{-11}$ and is only mildly affect by deviations from kinetic equilibrium.

Requesting that the QCD phase transition should not affect the freeze-out dynamics and that $\Delta N_{\rm eff}$ contributions for i2DM is in agreement with CMB data  be suppressed, we identify the cosmologically viable parameter region for i2DM candidates with mass and interaction strength in the range
\begin{align}
10\,\text{MeV} < m_1 < 1\,\text{GeV}\,,\quad  y >  10^{-11}.
\label{eq:parameters}
\end{align}
In this region, we have investigated possible effects of i2DM on Big Bang Nucleosynthesis, the Cosmic Microwave Background and supernova cooling, but find them much too small to modify current observations.

Laboratory searches for i2DM mostly rely on the production of dark
photons, subsequently decaying into dark fermions. Due to the feeble
interaction, the dark partner typically appears stable at the scales
of current colliders. Searches for mono-photons and missing momentum
at flavor experiments set a strong bound on the overall coupling of the
dark photon to the Standard Model, $\epsilon \lesssim 10^{-3}$. These bounds imply a minimum dark
fermion mixing and require efficient partner annihilation to satisfy
the relic abundance. The parameter range for viable i2DM is
thereby confined to light and compressed dark sectors with
\begin{align}
10\,\text{MeV} < m_1 < 200\,\text{MeV}\,,\quad \Delta < 0.1\,,\quad \epsilon < 10^{-3}\,.
\end{align}
At direct detection experiments, $\chi_1 N\to \chi_2 N$ up-scattering off atomic nuclei is not efficient enough
to further test i2DM, even if dark matter can be accelerated through
interactions with cosmic rays. On the other hand, long-lived dark partners can be produced in cosmic-ray up-scattering might leave an observable signature of $\chi_2 N\to \chi_2 N$ elastic scattering in the detector material. We leave this interesting avenue for future work.

Very promising probes of i2DM are fixed-target experiments with a long baseline, which can search for decays or scattering of long-lived dark partners in far detectors. Scattering is a prominent signal for light dark partners, while decays can be observed for heavier partners up to the GeV scale. Current searches for such signatures at CHARM and LSND already probe a significant portion of the i2DM parameter space, but lose steam for small mass splitting, where the momentum deposit in the detector is soft.

In the near future, neutrino experiments like ICARUS and SeaQuest or the beam-dump experiment LDMX can significantly improve the sensitivity to inelastic dark matter. Together with mono-photon searches at Belle II, they can conclusively test if inelastic Dirac dark matter is at the origin of the observed relic abundance.\\
\\

\begin{center} \textbf{Acknowledgments} \end{center}
\noindent We thank Alberto Mariotti for many interesting discussions
and for contributions at an early stage of this work. We also thank
Jan Heisig and Matteo Lucca for helpful discussions and Marco Hufnagel
for support with ACROPOLIS and exchanges.  The research of AF was
supported by the NWO Vidi grant ``Self-interacting asymmetric dark
matter". LLH is a research associate of the Fonds de la Recherche
Scientifique F.R.S.-FNRS. LLH and SJ have been supported by the FNRS
research grant number F.4520.19 and the IISN convention 4.4503.15. LLH
also acknowledges support of the ARC program of the Federation
Wallonie-Bruxelles. SJ is further supported by the Universit\'e Libre de Bruxelles PhD grant and the Strategic
Research Program High-Energy Physics and the Research Council of the
Vrije Universiteit Brussel. The research of SW has been supported by the German
Research Foundation (DFG) under grant no. 396021762--TRR 257.
%
\newpage
\appendix

\section{Formalism of i2DM}\label{app:model}
We provide further details about inelastic Dirac dark matter introduced in Sec.~\ref{sec:model}. The interactions of mass eigenstates described by the Lagrangian from Eq.~\eqref{eq:lagr} can be derived from gauge-invariant interactions
of the dark sector with the Standard Model in the unbroken phase of the underlying theory,
\begin{align}\label{eq:lagrangian}
	\mathcal{L} = & \ i\,\bar{\chi}_D \gamma^\mu D_\mu \chi_D -m_D \bar{\chi}_D \chi_D + i\, \bar{\chi}_0 \gamma^\mu \partial_\mu \chi_0 - m_0 \bar{\chi}_0 \chi_0\\\nonumber
	& \qquad + (D^\mu \phi_D)^\dagger (D_\mu \phi_D) + V(\phi_D) - (\lambda \phi_D \bar{\chi}_D \chi_0 + h.c.)\\\nonumber
	& \qquad - \frac{1}{4} \hat{F}_{D\mu\nu} \hat{F}_D^{\mu\nu} - \frac{1}{2} \frac{\epsilon}{c_W} \hat{B}_{\mu\nu} \hat{F}_D^{\mu\nu},
\end{align}
Here $D_\mu = \partial_\mu + i g_D Q_D\hat{A}_{D\mu}$ is the
covariant derivative for the gauge field $\hat{A}_D$ of the dark $U(1)_D$ symmetry. The field
strength tensors for the dark and the hypercharge gauge fields are
denoted as $\hat{F}_{D}^{\mu\nu}$ and $\hat{B}^{\mu \nu}$ respectively. The dark fermion field $\chi_D$ is charged under $U(1)_D$, while $\chi_0$ is a singlet under all gauge interactions.

The scalar $\phi_D$ with potential $V(\phi_D)$ and charge $Q_D=1$ facilitates spontaneous $U(1)_D$
breaking, once it develops a vacuum expectation value $v_D$, so that $\phi_D=(v_D+\varphi_D)/\sqrt2$.  Upon symmetry breaking, the dark photon $\hat{A}_D$ acquires a mass $m_{\hat{A}_D} = g_D v_D$, and the dark fermions $\chi_D$
and $\chi_0$ mix through the dark Yukawa coupling $\lambda$, resulting in two mass eigenstates
\begin{equation}
	\begin{pmatrix}
	\chi_1 \\ \chi_2
	\end{pmatrix}
	=
	\begin{pmatrix}
	\cos \theta & -\sin \theta \\
	\sin \theta & \cos \theta
	\end{pmatrix}
	\begin{pmatrix}
	\chi_0 \\ \chi_D
	\end{pmatrix},\qquad \sin(2 \theta) = \frac{\sqrt{2}\lambda v_D}{\Delta m_1}\,.
        \label{eq:rotferm}
\end{equation}
The corresponding masses are
\begin{align}
	m_{1,2} = \frac{1}{2}\big(m_D+m_0 \pm \Delta m_1\big) \quad {\rm with}\quad (\Delta m_1)^2 = (m_D-m_0)^2+2 (\lambda v_D)^2\,,
\end{align}
where $\Delta$ is the relative mass difference defined in Eq.~(\ref{eq:Delta}).
In order to obtain canonical kinetic terms in Eq.~\eqref{eq:lagrangian},
we first redefine the $U(1)_Y$ and $U(1)_D$ gauge bosons to absorb the kinetic mixing via
\begin{equation}
	\begin{pmatrix}
	\hat{B}^\mu  \\ \hat{A}^\mu_D
	\end{pmatrix}
	= G(\hat{\epsilon})
	\begin{pmatrix}
	\tilde{B}^\mu \\ \tilde{A}^\mu_D
	\end{pmatrix}
	=
	\begin{pmatrix}
	1  && -\frac{\hat{\epsilon}}{\sqrt{1-\hat{\epsilon}^2}}\\
	0  && \frac{1}{\sqrt{1-\hat{\epsilon}^2}}
	\end{pmatrix}
	\begin{pmatrix}
	\tilde{B}^\mu  \\ \tilde{A}^\mu_D
	\end{pmatrix},
        \label{eq:trsfkin}
\end{equation}
where $\hat{\epsilon}=\epsilon/\cos \theta_W$. Subsequently, we transform the fields to the physical eigenstates of weak interactions, $A^\mu,\,Z^\mu,\,A'^{\mu}$, via two rotations:
one rotation, $R_\xi$, mixing the $SU(2)$ gauge field $W^3_\mu$
and $\tilde{A}^\mu_D$ with an angle $\xi$; and a second rotation,
$R_W$, mixing $W^3_\mu$ and $\tilde{B}_\mu$ with the Weinberg angle
$\theta_W$. The overall transformation takes the form
\begin{equation}
\begin{pmatrix}
\tilde B^\mu \\ W^{3\mu} \\ \tilde A^\mu_D
\end{pmatrix}
=
\begin{pmatrix}
c_W && -c_\xi s_W && s_\xi s_W  \\
s_W && c_\xi c_W && -s_\xi c_W \\
0 && s_\xi && c_\xi
\end{pmatrix}\,
\begin{pmatrix}
A^\mu \\ Z^{\mu} \\ A'^\mu
\end{pmatrix},
\label{eq:rotgauge}
\end{equation}
where
\begin{eqnarray}
  &&\tan 2 \xi = \frac{2 \eta s_W}{1-(\eta s_W)^2-\delta} \label{eq:xi}\\\nonumber
  &&{\rm with}\quad \eta=\hat{\epsilon} / \sqrt{1-\hat{\epsilon}^2}, \quad \delta =\hat{\delta}/(1-\hat{\epsilon}^2) ,\quad  \hat{\delta}=m_{\hat{A}_D}^2/m_{\hat{Z}}^2\,,
\end{eqnarray}
and $m_{\hat{Z}}= g v/(2c_W)$, with the $SU(2)$ gauge coupling, $g$, and the SM Higgs vacuum expectation value, $v$. The masses of the $Z$ boson and the dark photon are
\begin{eqnarray}
    &&m^2_Z=m_{\hat{Z}}^2\left((c_\xi+\eta s_\xi s_W)^2+s_\xi^2\delta\right)\quad {\rm and}\quad m^2_{A'}=m_{\hat{Z}}^2\left((- s_\xi + c_\xi\eta s_W)^2+c_\xi^2\delta\right),
  \label{eq:gbmasses}
\end{eqnarray}
while the photon remains massless. In the limit of small kinetic
 mixing, $\{\epsilon, \eta,\xi \} \ll 1$, one
recovers $ m_Z\simeq m_{\hat{Z}}$ and $m_{A'}\simeq
m_{\hat{A}_D}$~\cite{Wells:2008xg,
  Curtin:2014cca,Freitas:2015hsa,Foldenauer:2019vgn}.

 To derive the Feynman rules and transition amplitudes, we use {\tt
  FeynRules}~\cite{Alloul:2013bka}. In particular, we have adapted the
\emph{Hidden Abelian Higgs Model}~\cite{Wells:2008xg}, which is available from the
{\tt FeynRules} model data base to i2DM. Our
 {\tt FeynRules} model for i2DM is available at~\cite{FeynRulesModel}. In terms of mass eigenstates, the Lagrangian finally reads
\begin{align}\label{eq:lag-mass}
	\mathcal{L} \supset & \ -e \big(A_\mu - \epsilon A'_{\mu}\big) J^\mu_{em} + \frac{e}{2 s_W c_W} Z_\mu J^\mu_Z\\\nonumber
	& - g_D \left(A'_{\mu} + \epsilon\, t_W Z_\mu\right) \big(s_\theta^2  J^\mu_1 + c_\theta^2 J^\mu_2 - c_\theta s_\theta J^\mu_{12}\big) \\ \nonumber
				& - \frac{\lambda}{\sqrt{2}} \varphi_D\big(c_{2\theta} (\bar{\chi}_1 \chi_2 + \bar{\chi}_2 \chi_1) - s_{2\theta} (\bar{\chi}_1 \chi_1 - \bar{\chi}_2 \chi_2)\big)\\ \nonumber
			&+\frac{g_D^2}{2} \big( 2 v_D \varphi_D + \varphi_D^2\big) \big(A'_{ \mu} A'^\mu + \epsilon\,t_W(A'_{ \mu} Z^\mu+Z_\mu A'^\mu) + \epsilon^2 t_W^2 Z_\mu Z^\mu \big)\,,
\end{align}
where $t_W$ refers to the tangent of the Weinberg angle $\theta_W$.
The dark currents $J^\mu_i$ with $(i=1,2)$ and $J^\mu_{12}$ are
defined in Eq.~(\ref{eq:J12}). The SM fermion currents are given by
\begin{align}
J^\mu_Z = \bar{f}(c_V \gamma^\mu - c_A \gamma^\mu \gamma^5)f,\quad J^\mu_{em} = Q_f \bar{f} \gamma^\mu f,
\end{align}
where $f$ are the SM fermions and $c_V=T_f^3 - 2s_W^2 Q_f$, $c_A= T_f^3$ are the electroweak charges, with $T_f^3$ the weak isospin quantum number and $Q_f$ the electric charge in units of $e$. Here we have only included the leading terms in $\epsilon$ and provided the SM fermion couplings in unitary gauge.

From the transformations of Eqs.~(\ref{eq:trsfkin}) and
(\ref{eq:rotgauge}) and the Lagrangian in Eq.~\eqref{eq:lag-mass}, it should be clear that the photon $A^\mu$ has
no field component from the dark $U(1)_D$ gauge field
$\hat{A}^\mu_D$. As a result, the photon has no fundamental couplings to the dark fermions $\chi_1,\,\chi_2$. In particular, the dark matter candidate $\chi_1$ carries no millicharge and is not subject to otherwise strong constraints~\cite{Hambye:2019dwd}.

\section{Boltzmann equations beyond thermal equilibrium}
\label{app:CollisionTerm}
On general grounds, in an isotropic and homogeneous universe the evolution of a particle species $i$ is described in terms of
a distribution function $f_i(t,|\vec p_i|)$,
 expressed in terms of the physical time $t$ and of the norm of the physical 3-momentum $\vec p_i$, denoted as $|\vec p_i| \equiv p_i$.\footnote{In the text, $p_i$ usually refers to the norm of the
  3-momentum, except when it appears in a 4-dimensional delta
  function which enforces both 3-momentum {\it and} energy conservation.}
  If a species $i$ interacts with other species $j$, the time evolution of species $i$ is described by the Boltzmann equation
\begin{equation}
  \frac{d f_i(t,p_i)}{dt}=\frac{1}{E_i}\, {\cal C}[f_i(t,p_i),f_j(t,p_j)]\,,
  \label{eq:UnintBE0}
\end{equation}
where ${\cal C}[f_i(t,p_i),f_j(t,p_j)]$ is the collision term
involving all decay and scattering processes with the other species $j$, and
$E_i=\sqrt{ p_i^2+m_i^2}$ is the energy of particles of species $i$ with mass $m_i$. In this appendix, and in particular in Sec.~\ref{sec:kin-dec}, we give more details about the collision terms for coscattering, ${\cal C}_{\rm coscat}$, and 3-body
inverse decays, ${\cal C}_{\rm decay}$.

The total time derivative in
Eq.~(\ref{eq:UnintBE0}) can be re-expressed in terms of partial
derivatives with respect to time and momentum as
\begin{eqnarray}
    \frac{d f_i(t,p_i)}{dt}&=&\left( \partial_t - H p_i \partial_{p_i}
    \right) f_i(t,p_i) =\overline{H} \left(x\partial_x+ \frac{x}{ 3
      h_{\rm eff}}\frac{d h_{\rm eff}}{dx} \, q_i\,\partial_{q_i} \right) f_i(x,q_i),
  \label{eq:UnintBE1}
\end{eqnarray}
with
\begin{equation}
    \overline{H} = H\left(1-\frac{x}{3 h_{\rm eff}}\frac{dh_{\rm eff}}{dx}\right)^{-1}.
\label{eq:barH}
\end{equation}
In Eq.~(\ref{eq:UnintBE1}) we have used the rescaled time $x$ and momentum
variable $q_i$ introduced in Eq.~(\ref{eq:xq}).  In this appendix,
we will make use of the variables $(t,p_i)$ or $(x,q_i)$ whenever
convenient. Also, $H=H(x)$ is the Hubble expansion rate and $h_{\rm
  eff}= h_{\rm eff}(x)$ is the effective number of relativistic
degrees of freedom contributing to the entropy density $s=h_{\rm
  eff}\, {2\pi^2}/45 \, T^3$.\footnote{The insertions of $d h_{\rm
    eff}/dx$ in Eqs.~(\ref{eq:UnintBE1})
  and~(\ref{eq:barH}) are due to the usual choice of rescaled momentum
  $q= p/T$, instead of the time-independent comoving momentum $q=p a$, where $a$ is the scale factor. Another convenient
  choice of time-independent rescaled momentum would be
  $q=p/s^{1/3}$ see e.g.~\cite{Belanger:2020npe,Decant:2021mhj}.} In
this paper we consider dark matter production in a radiation dominated
era, in which case the Hubble rate reduces to
\begin{eqnarray}
 H(x)&=&\frac{m_1^2}{x^{2}M_0}  \quad {\rm with}\quad  M_0=M_P \sqrt{\frac{45}{4 \pi^3 g_*}}\,,
  \label{eq:H}
\end{eqnarray}
where $M_P$ is the Planck mass and $g_\ast=g_\ast(x)$ denotes the number of
relativistic degrees of freedom  in the  thermal bath at time $x$
contributing to the radiation energy density $\rho_R= g_\ast\, \pi^2/30\,
T^4$.  For our numerical analysis, we use tables available in the public code {\tt micrOMEGAs}~\cite{Belanger:2018ccd} to evaluate $h_{\rm eff}(x)$ and  $g_\ast(x)$. When the number of relativistic
degrees of freedom, $h_{\rm eff}$ and $g_\ast$, can be considered
constant around freeze-out, the unintegrated
Boltzmann equations of Eq.~(\ref{eq:UnintBE1}) simplify to
 \begin{equation}
    x H \, \partial_x f_i(x,q_i)=\frac{1}{E_i} {\cal C}[f_i(x,q_i),f_j(x,q_j)]\,,
      \label{eq:UnintBE3}
\end{equation}
as reported in Eq.~(\ref{eq:UnintBE}).  To compute the dark matter freeze-out beyond
kinetic and chemical equilibrium in i2DM, we will use the latter evolution equation, see
Sec.~\ref{sec:kin-dec} for details.

In what follows, we will encounter the equilibrium number densities $n^i_{\text{eq}}(t)$, i.e., the number densities obtained by integrating over the kinetic equilibrium distribution functions $f^{\rm eq}_i(t,p_i)$, where $f^{\rm eq}_i(t,p_i)$ are the Fermi-Dirac, Bose-Einstein or Boltzmann distributions for zero chemical potential.

\subsection{Deviation from chemical equilibrium before freeze-out}
\label{sec:chem-dec}
%
  In i2DM, as long as deviations of $\chi_1$ from kinetic equilibrium with the heat
 bath can be neglected up until freeze-out, the evolution of $\chi_{1,2}$ can be described by
the number densities $n_{1,2}(x)$, or equivalently by the
 comoving number densities $Y_{1,2}(x)$. Within this framework, it is
 useful to introduce the reaction densities for
 decays and scattering,
\begin{eqnarray}\label{eq:reaction-densities}
 \gamma_{i\to j k l}&=&\int d\phi_i \phi_j d\phi_k d\phi_l  f_i^{\rm eq}(t,p_i) (2\pi)^4\delta^4(p_i-p_j-p_k-p_l) |{\cal M}_{i\to j kl}|^2  \label{eq:gamm}\\
     \gamma_{ij\to kl}&=&\int d\phi_i d\phi_j f_i^{\rm eq}(t,p_i) f_j^{\rm eq}(t,p_j) \int  d\phi_k d\phi_l (2\pi)^4\delta^4(p_i+p_j-p_k-p_l) |{\cal M}_{ij\to kl}|^2\,,\nonumber
\end{eqnarray}
which can be rewritten in terms of the thermally averaged cross section
and the decay rate in the rest frame of species $i$, as in
Eq.~(\ref{eq:gam-min}). In Eq.~(\ref{eq:gamm}), $|{\cal M}|^2$ is
the squared scattering amplitude, averaged over initial- {\it and}
final-state degrees of freedom, and $d\phi_i=g_i d^3p_i/(2E_i(2\pi)^3)$ denotes
the phase-space element.

As long as $\chi_{1,2}$ can be assumed to be in
kinetic \emph{and} chemical equilibrium before freeze-out, the set of partial integro-differential equations for the dark fermions from Eq.~(\ref{eq:UnintBE0}), integrated over the 3-momenta $p_i$, reduces to
 one single well-known ordinary differential equation for the total dark sector number density $n(t)= \sum_{i=1,2}
n_i(t)$~\cite{Griest:1990kh,Gondolo:1990dk}. This simplification
applies to freeze-out in the phases of coannihilation and partner
annihilation in region (A). In this regime,
one can compute the dark matter relic abundance with any publicly available Boltzmann solver, such
as {\tt micrOMEGAs}~\cite{Belanger:2018ccd},
{\tt DarkSUSY}~\cite{Bringmann:2018lay} or {\tt MadDM}~\cite{Ambrogi:2018jqj}.

In the conversion phase, the very small interaction strength $y$
implies strongly suppressed coscatterings and decays
 and chemical equilibrium between $\chi_1$ and $\chi_2$ is lost. As a result,
the relation $n_1/n_2=n_{1}^{\rm eq}/n_{2}^{\rm eq}$ cannot be
assumed. In order to take deviations from chemical equilibrium into account, we have
to solve a coupled system of Boltzmann equations
including the conversion rates $\gamma_{1\to 2}$~\cite{Garny:2017rxs,Garny:2018icg,DAgnolo:2018wcn,Junius:2019dci,Herms:2021fql},\footnote{Similar Boltzmann equations exist for the antiparticles $\bar{\chi}_1$ and $\bar{\chi}_2$. In i2DM these equations are completely equivalent and we  obtain the total abundance for $\chi_1+\bar{\chi}_1$ by simply doubling the yield for $\chi_1$.}
\begin{align}
\label{eq:BE1}
\frac{dY_1}{dx}=\frac{-1}{\overline{H} xs} & \left[\gamma_{1 1} \left(\frac{Y^2_{1}}{{Y_{1}^{\rm eq}}^2}-1 \right) + \gamma_{1 2} \left(\frac{Y_{1} Y_{2}}{Y_{1}^{\rm eq} Y_{2}^{\rm eq}}-1 \right) -   \gamma_{2   \to 1 }\left(  \frac{Y_{2}}{Y_{2}^{\rm eq}}-\frac{Y_{1}}{Y_{1}^{\rm eq}} \right)  \right]\\\nonumber
\frac{dY_{2}}{dx}=\frac{-1}{ \overline{H}xs} & \left[\gamma_{22} \left( \frac{Y^2_{2}}{{Y_{2}^{\rm eq}}^2}- 1\right) + \gamma_{12} \left(\frac{Y_{1} Y_{2}}{Y_{1}^{\rm eq} Y_{2}^{\rm eq}}-1 \right)  +\gamma_{2\to1}\left(  \frac{Y_{2}}{Y_{2}^{\rm eq}}-\frac{Y_{1}}{Y_{1}^{\rm eq}} \right) \right],
\end{align}
where $\gamma_{ij}$ are the reaction densities defined in
Eq.~\eqref{eq:gamij} and $\gamma_{i\to j}= \gamma_{i\to
  j}^{dec}+\gamma_{i\to j}^{scat}$.
For i2DM, we have checked that we can neglect quantum statistical effects in the relevant density distributions $f_i$ and use
the Maxwell-Boltzmann equilibrium distributions $f_i^{\rm
  eq}$.  In i2DM, Eq.~\eqref{eq:BE1} can safely be used for freeze-out scenarios
   in regions (A) and (B), where kinetic equilibrium is maintained up to freeze-out.\footnote{For the scenarios considered in this paper, the results we obtain by
  carefully accounting for kinetic decoupling, see
  Sec.~\ref{sec:kin-dec}, are very close to those
  derived using Eq.~(\ref{eq:BE1}), see the
  discussion around Fig.~\ref{fig:ratiogamH}.}

\subsection{Deviation from chemical and kinetic equilibrium around freeze-out}
\label{sec:kin-dec}
%
When the conversion rate is around 10 to 100 times the Hubble rate at the time of $\chi_2$ chemical
decoupling, dark matter can no longer be expected to be kept in
chemical or kinetic equilibrium with $\chi_2$ around the DM freeze-out time $x_{\rm fo}$. The assumptions for Eq.~(\ref{eq:BE1}) do not hold anymore, and a priori one needs to solve Eq.~(\ref{eq:UnintBE3}) including the full momentum and time dependentce in the $\chi_1$ distribution function $f_1(x,q_1)$.

For i2DM, we apply two levels of simplification to Eq.~(\ref{eq:UnintBE3}). First, we neglect the time dependence in $h_{\rm eff}$ for the dark matter density evolution, setting $h_{\rm eff}=h_{\rm eff}(x_{\rm fo})$ when integrating over $f_1(x,q_1)$. For dark fermions in the MeV-GeV mass range, we obtain a good
approximation to the exact solution of Eq.~(\ref{eq:UnintBE3}). In this mass range, freeze-out is expected to happen between the QCD phase transition and neutrino decoupling, a period in which $h_{\rm eff}\simeq 10.75$ is
  fairly constant.  Second, we follow Refs.~\cite{DAgnolo:2017dbv,
  Garny:2017rxs} and only take into account the dominant interaction processes
 driving the dark matter distribution $f_1(x,q_1)$ towards kinetic
equilibrium. For i2DM, this means that we include conversion processes but neglect coannihilation.

With these simplifications, the unintegrated Boltzmann equation for
$f_1(x,q_1)$ from Eq.~(\ref{eq:UnintBE3}) can be rewritten as
\begin{eqnarray}
  H x\partial_x f_1(x,q)&=& \tilde C_{1\to 2}(x,q) \left(f_1^{\rm eq}(x,q) \frac{Y_2(x)}{Y_2^{\rm eq}(x)}-f_1 (x,q)\right),
  \label{eq:UBcoscat}
\end{eqnarray}
where $q=q_1 = p_1/T$ to simplify the notation. The
contributions to the collision operator $\tilde C_{1\to 2}(x,q)=\tilde
C_{\rm coscat}(x,q)+\tilde C_{\rm decay}(x,q) $ from coscattering and
decays are spelled out in Secs.~\ref{sec:coscattering}
and~\ref{sec:three-body-decay}. The above description assumes that
$\chi_2$ and the light SM fermions involved in the conversion
processes are in kinetic equilibrium.  We also neglect all spin
statistics effect.

The differential equation in Eq.~(\ref{eq:UBcoscat}) can be solved iteratively with $Y_2(x)$ as an input. The latter is obtained
from the integrated Boltzmann equation for $Y_2$ in Eq.~(\ref{eq:BE1}), which in
turn involves $Y_1(x)$, or equivalently the zeroth moment of $f_1(x,q)$ in $q$,
 obtained by integrating over Eq.~(\ref{eq:UBcoscat}). More details on
the integration of Eq.~(\ref{eq:UBcoscat}) will be discussed in
Sec.~\ref{sec:simpl-unint-boltzm}.

\subsubsection{Coscattering}
\label{sec:coscattering}
Here we provide details on the collision term due to coscattering. First we show that the collision term for coscatterings $\chi_1 f \rightarrow \chi_2 f'$ can indeed be written as in
Eq.~\eqref{eq:UBcoscat}. Neglecting spin-statistic effects, the
collision term takes the form
\begin{align}
  {\cal C}_{\text{coscat}} = \frac{1}{2}\int d\phi_f d\phi_2 d\phi_{f'} (2\pi)^4 \delta^{(4)}(p_1+p_f-p_2-p'_{f}) |{\cal M}_{\chi_1
f \rightarrow \chi_2 f'} |^2  \left( f_2 f_{f'} - f_1 f_{f} \right).
\end{align}
Assuming that the SM fermions $f$ and $f'$ are in
chemical and kinetic equilibrium with the thermal bath and that $\chi_2$
is in kinetic equilibrium throughout the $\chi_1$ freeze-out, we have $f_{x}(t,p_{k})=f^{\rm eq}_{k}
(t,p_{k})$ for $k=\{f,f'\}$ and $f_2(t,p_2) = f_2^{\rm
  eq}(t,p_2) \, Y_2(t)/Y_2^{\rm eq}(t)$, so that
  \begin{equation}
    f_2 f_{f'} - f_1 f_{f} = \left( f_2^{\rm eq} \frac{Y_2}{Y_2^{\rm eq}} f_{f'}^{\rm eq} - f_1 f_{f}^{\rm eq} \right) = f_f^{\rm eq} \left( f_1^{\rm eq} \frac{Y_2}{Y_2^{\rm eq}}  - f_1 \right).\label{eq:distrib-rel1}
\end{equation}
In the second equation we have used the relation of detailed balance,
$f_2^{\rm eq} f_{f'}^{\rm eq} = f_1^{\rm eq} f_f^{\rm eq}$.
With  Eq.~(\ref{eq:distrib-rel1}), we can write the collision term as
\begin{equation}
\label{eq:CollTermCoscat}
    \frac{1}{E_1}\, {\cal C}_{\text{coscat}} = \tilde {\cal C}_{\text{coscat}} \left( f_1^{\rm eq} \frac{Y_2}{Y_2^{\rm eq}}  - f_1 \right),
\end{equation}
with the coscattering collision operator
\begin{eqnarray}
    \tilde {\cal C}_{\text{coscat}}     &=&\frac{g_f T}{ 16 \pi^2 p_1E_1}\int ds\, \sigma_{\text{coscat}}(s)\, (s-m_1^2) \left(e^{-E_{f}^-/T}- e^{-E_{f}^+/T}\right).
\label{eq:Ctildcoscatfin}
\end{eqnarray}
Here the cross section for coscattering process is defined as~\cite{DAgnolo:2018wcn,Garny:2018icg}
 \begin{equation}
     \sigma_{\text{coscat}}(s) = \frac{1}{4 \hat{p}_1 \sqrt{s}} \int d\phi_2 d\phi_{f'} \ (2\pi)^4 \delta^4(p_1+p_f-p_2-p'_{f}) \ |{\cal M}_{\chi_1
f \rightarrow \chi_2 f'} |^2 \,,
 \end{equation}
 with the modulus of the dark matter 3-momentum in the
 centre-of-mass frame, $\hat{p}_1$, the squared center-of-mass energy, $s$, and the energy variables
 \begin{equation}
     E_f^\pm = \frac{s-m_1^2}{2m_1^2} (E_1 \pm p_1)\,.
 \end{equation}

\subsubsection{Three-body decays}
\label{sec:three-body-decay}
As mentioned in Sec.~\ref{sec:relevant-processes}, the inverse decay
process $ \chi_1 f f'\to \chi_2 $ can play a role in keeping $\chi_1$
in kinetic equilibrium. The corresponding collision reads
\begin{eqnarray}
  {\cal C}_{\text{decay}} = \frac{1}{2}\int\! d\phi_2 d\phi_f d\phi_{f'} (2\pi)^4 \delta^{(4)}(p_2-p_1-p_f-p'_{f}) |{\cal M}_{\chi_2 \to \chi_1 f f'}|^2  \left( f_2 - f_1 f_{f} f_{f'} \right).
\end{eqnarray}
Assuming again that all involved SM fermions $f,f'$ are in chemical and kinetic
equilibrium with the thermal bath and that $\chi_2$ is in kinetic
equilibrium with the bath, the collision term reduces to
\begin{equation}
\label{eq:CollTermDecay}
    \frac{1}{ E_1}{\cal C}_{\text{decay}} = \tilde {\cal C}_{\text{decay}} \left( f_1^{\rm eq} \frac{n_2}{n_2^{\rm eq}}  - f_1 \right),
\end{equation}
with the collision operator for decays,
\begin{eqnarray}
    \tilde {\cal C}_{\text{decay}}
    &=& \frac{g_{f'}}{2 E_1}\int d\phi_2 d\phi_f \frac{2\pi}{2E'_{f}} \delta(E_2-E_1-E_f-E'_{f}) |{\cal M}_{\chi_2 \to \chi_1 f f'}|^2  f_{f}^{\rm eq} f_{f'}^{\rm eq}
    \label{eq:CtildeInterediate1}
\end{eqnarray}
after integrating over $d\phi_{f'}$.

Applying the methodology from~\cite{DEramo:2020gpr} for 3-body
decays and neglecting the SM fermion masses, the
collision operator can be rewritten as
\begin{align}\label{eq:Ctilddecfin}
    \tilde {\cal C}_{\text{decay}} = \frac{g_f g_{f'} g_2}{256 \pi^3 p_1 E_1} & \int_0^{(\Delta m_1)^2}\!\!\!\! \frac{dm_{ff'}^2}{\sqrt{\lambda(m_{ff'},m_2,m_1)}}\\\nonumber
    & \times \int_{E_2^-}^{E_2^+} dE_2 \,f_{f}^{\rm eq} f_{f'}^{\rm eq}   \int_{(m_{1f}^2)^-}^{(m_{1f}^2)^+} dm_{1f}^2 |{\cal M}_{\chi_2 \to \chi_1 f f'}|^2,
\end{align}
where
\begin{eqnarray}
 \lambda(x,y,z) & = & (x^2-(y+z)^2)(x^2-(y-z)^2),\\\nonumber
    m_{1f}^2 &=& (p_1+p_f)^2,\\\nonumber
   m_{ff'}^2 &=& (p_2-p_1)^2.
\end{eqnarray}
The integration boundaries are
\begin{eqnarray}
    &E_2^\pm& = \sqrt{m_2^2+(p_2^\pm)^2},\quad {\rm with} \\\nonumber
    &p_2^\pm& = \frac{p_1(m_1^2+m_2^2-m_{ff'}^2)\pm \sqrt{(m_{ff'}^2+m_1^2)\lambda(m_{ff'},m_2,m_1)}}{2m_1^2},
\end{eqnarray}
and $(m_{1f}^2)^\pm = m_1^2 + m_{ff'} (\hat{E}_1\pm\hat{p}_1)$, where the hatted quantities are evaluated in the reference frame where $\Vec{p}_f = \Vec{p}_{f'}$.

\subsubsection{Simplified unintegrated Boltzmann equation}
\label{sec:simpl-unint-boltzm}
The collision terms for inverse decays
and coscattering from Eqs.~\eqref{eq:CollTermCoscat} and
\eqref{eq:CollTermDecay} have the same form as the right-hand side
of the simplified Boltzmann equation in Eq.~(\ref{eq:UBcoscat}). We can solve the latter for
$\tilde{\cal C}_{1\to2} = \tilde{\cal C}_{\text{coscat}} + \tilde{\cal
  C}_{\text{decay}}$ using the results of
Eqs.~(\ref{eq:Ctildcoscatfin}) and~(\ref{eq:Ctilddecfin}) and obtain
\begin{eqnarray}
\label{eq:UnintBEsimp}
    \frac{\partial f_1(x,q)}{\partial x} + g(x,q)f_1(x,q) = g(x,q) h(x,q),
\end{eqnarray}
where
\begin{eqnarray}
    g(x,q) &=& \frac{\tilde{\cal C}_{2\to1}(x,q)}{xH(x)},\\\nonumber
    h(x,q) &=& f_1^{\rm eq}(x,q)\frac{Y_2(x)}{Y_2^{\rm eq}(x)}.
\end{eqnarray}
Multiplying both sides of Eq.~(\ref{eq:UnintBEsimp}) by $u(x,q)=\exp \left[ \int dx \, g(x,q) \right]$ and using $\frac{\partial u(x,q)}{\partial x} = g(x,q)u(x,q)$, we can simplify this equation to
\begin{eqnarray}
    \frac{\partial (u(x,q)f_1(x,q))}{\partial x} = \frac{\partial u(x,q)}{\partial x} h(x,q).
\end{eqnarray}
Integrated by parts, we obtain
\begin{eqnarray}
    u(x,q)f_1(x,q)=u(x_0,q)f_1(x_0,q)+\left[h(x,q) u(x,q) \right]^x_{x_0}+\int_{x_0}^x dz \frac{\partial h(z,q)}{\partial z} u(z,q)\,,
\end{eqnarray}
where we set the initial time to $x_0=1$.

In i2DM, at early times we can use the
 boundary conditions that $\chi_1$ is in kinetic equilibrium, $f_1(x_0,q) = f_1^{\rm eq}(x_0,q)$, and that $\chi_2$ is in chemical equilibrium, $Y_2(x_0)=Y_2^{\rm eq}(x_0)$, such that $h(x_0,q)=f_1(x_0,q)$. The dark matter phase-space distribution then reduces to
\begin{align}\label{eq:f1-simp}
    f_1(x,q) & = f_1^{\rm eq}(x,q)\frac{Y_2(x)}{Y_2^{\rm eq}(x)}\\\nonumber
    & \quad - \int_{x_0}^x dz \frac{\partial }{\partial z}\left( f_1^{\rm eq}(z,q)\frac{Y_2(z)}{Y_2^{\rm eq}(z)} \right) \exp\left(\int_z^x dy \frac{\tilde{\cal C}_{2\to1}(y,q)}{yH(y)}\right).
\end{align}
In order to solve this equation, we have to specify the comoving
number density of $\chi_2$, which remains in kinetic equilibrium throughout
the whole evolution of $\chi_1$. Hence, we can use the integrated Boltzmann equation
for $\chi_2$ from Eq.~(\ref{eq:BE1}).

Solving Eqs.~\eqref{eq:f1-simp} and~(\ref{eq:BE1}) together is numerically
difficult, so we choose to solve the two equations iteratively. In a first step, we solve the system of integrated
Boltzmann equations from Eq.~(\ref{eq:BE1}). This gives
us an initial value for $Y_2(x)$. We then solve the unintegrated Boltzmann
Eq.~(\ref{eq:UnintBEsimp}) for $\chi_1$ and feed again the
integrated Boltzmann equation for $\chi_2$ from Eq.~(\ref{eq:BE1}) to obtain the next iteration for
$Y_2(x)$. We stop this iteration once the difference in the relic dark matter abundance between the last two iterations is less
than one percent.


\newpage
\bibliographystyle{JHEP}
\bibliography{inelastic-dirac-dark-matter}


\end{document}